\documentclass[]{article}
\usepackage[utf8]{inputenc}
\usepackage{authblk}
\usepackage{amsmath}
\usepackage{amssymb}
\usepackage{graphicx}
\usepackage{setspace}
\usepackage{xcolor}
\usepackage[margin=1in]{geometry}

\usepackage[backend=biber,style=numeric,sorting=none]{biblatex}
\bibliography{pmaca1refs.bib}
\title{Physical mechanisms affecting critical angle for nanopatterning in irradiated thin films: II. Collision cascade details}
\author{Tyler Evans \& Scott Norris}
\affil{Department of Mathematics, Southern Methodist University, Dallas, TX 75275, United States of America}

\begin{document}

\maketitle

\begin{abstract}
Ion-beam irradiation of an amorphizable material such as Si or Ge may lead to spontaneous pattern formation beyond some critical angle of the beam versus the surface. It is known from experimental results that this critical angle varies according to beam energy, ion species and target material. However, most prevailing theoretical analyses predict a critical angle of $45^{\circ}$ independent of energy, ion and target, disagreeing with experiment. In this second part of a set of papers, we consider the influence of the relationship between the upper and lower interfaces of the amorphous thin film (the ``interface relation"). From our previous work, we are motivated to derive from a geometric argument closed-form expressions describing the interface relation in terms of the collision cascade shape. This feature leads to a refined characterization of the influence of ion-, target- and energy-dependence on critical angle selection. Then, with an estimate from experimental data of the allocation of beam energy to each of isotropic and deviatoric strain, we seek to connect the experimentally-observed $\theta_c \approx 45^{\circ}$ to a purely mechanical theory for the first time. Intriguingly, we find that the hypothesis that deviatoric and isotropic stress components occur with the same distribution throughout the bulk cannot explain the data. Error-minimization between experimental data and model predictions for in-plane stress leads to a physically-interesting modification to our original hypothesis appears to lead immediately to parameter estimates that yield good agreement between theory and experimental observations of critical angle in the 250eV Ar$^+ \to$ Si system. This new hypothesis also appears to provide good qualitative agreement across two different particle-substrate systems and a wide range of energies and is sufficient to suggest the physical origin of the phenomenological mechanisms used throughout the low-energy irradiation literature. Our findings immediately prompt further experimental studies of isotropic swelling and angle-dependent stress evolution in irradiated thin films for different materials and energy levels, as well as further theoretical study of the physical origin of the deviatoric and isotropic stress components in irradiated, amorphizable targets.
\end{abstract}

\tableofcontents


\section{Introduction}
Experiments as early as about the 1960's have observed that the energetic irradiation of a substrate by an ion-beam may lead to the spontaneous self-organization of patterns and structures, ranging from hexagonal arrays of dots, to parallel or perpendicular mode ripples, terraces, conical structures, and more, sometimes with characteristic lengths on the scale of only a few nanometers \cite{navez-etal-1962,ziberi-etal-APL-2008,madi-etal-2008-PRL,madi-etal-PRL-2011,madi-aziz-ASS-2012}. The specific patterns that form are known to vary according to many parameters of the system, including, but by no means limited to, the projectile, beam energy, beam angle, target species, ambient temperature, chemical reactivity of projectile and target, whether the target becomes amorphous upon sustained exposure \cite{NorrisAziz_predictivemodel}. These early experiments were well-timed relative to the advent of the computing, and the potential to mass-manufacture useful nano-engineered materials by simply irradiating a large sheet of elemental semiconductor material and exploiting their properties of self-organization was soon recognized. Such an approach would lead to cheap, fast, scalable nano-scale manufacturing and avoid the problems associated with attempting to directly manipulate matter at the atomic scale, favoring instead the tuning of the macroscale parameters determining the nano-scale self-organization to steer the system dynamics towards the desired configuration.

This potential has clearly not been realized, and nano-scale engineering continues to be relatively slow and difficult. Many challenges exist to full macroscale control of ion-induced, self-organized nano-structure formation, especially a lack of a unified physical theory. Even in recent years, there is no consensus as to the exact physics of self-organization, and proposals of possible new mechanisms to explain experimentally-observed behaviors are varied and ongoing. However, some consensus has been built about broad details, sometimes with striking qualitative and near-quantitative accuracy \cite{chan-chason-JAP-2007,munoz-garcia-etal-MSER-2014,NorrisAziz_predictivemodel} without necessarily suggesting the underlying physics.

Most modeling approaches focus either on the prompt regime, occurring at the time-scale of $\sim10^{-9}$ seconds, or the gradual regime, occurring at the time-scale of $\sim10^{2}$ seconds, with very little work having been done to consider both time-scales simultaneously. Historically, work in the prompt regime has considered the effects of the ion-beam directly on the surface, especially erosion \cite{sigmund-PR-1969,sigmund-JMS-1973,bradley-harper-JVST-1988}, lateral redistribution \cite{carter-vishnyakov-PRB-1996} and surface diffusion \cite{makeev-etal-NIMB-2002,cuerno-barabasi-PRL-1995}. Work in the gradual regime has tended to consider the near-surface region of the irradiated substrate, including stress evolution \cite{castro-cuerno-ASS-2012,norris-PRB-2012-viscoelastic-normal}, defect dynamics and their interactions with stress \cite{chan-chason-JVSTA-2008,ishii-etal-JMR-2014}, the weakening of molecular bonds \cite{wesch-wendler-book-2016}, melt-cycles \cite{wesch-wendler-book-2016,van-dillen-etal-APL-2001-colloidal-ellipsoids,van-dillen-etal-APL-2003-colloidal-ellipsoids,van-dillen-etal-PRB-2005-viscoelastic-model}, and, in recent years, flow within the amorphous bulk, variously characterized by effective body-forces \cite{castro-cuerno-ASS-2012,castro-etal-PRB-2012,moreno-barrado-etal-PRB-2015,munoz-garcia-etal-PRB-2019} or anisotropic plastic flow. 

Recently, an apparent mismatch between the prompt-regime theory and experiment was considered \cite{umbach-etal-PRL-2001}, with surface relaxation evidently occurring without temperature dependence, despite surface diffusion, as proposed, being a fundamentally thermally-driven relaxation mechanism. Observing that in some materials the ongoing exposure to ion-bombardment leads to the development of a heavily-damaged amorphous region near the surface \cite{wesch-wendler-book-2016}, it was proposed \cite{umbach-etal-PRL-2001} that theory and experiment could be reconciled by exchanging thermally-driven surface diffusion for viscous relaxation of the heavily-damaged near-surface region, effectively treating the amorphous region as a highly-viscous thin film. The scaling of surface relaxation with wavenumber according to a surface-confined viscous flow model was later demonstrated in an experimental-theoretical collaboration \cite{norris-etal-SREP-2017}, verifying the need to consider gradual-regime mechanisms alongside prompt-regime ones in at least some cases. It has also been shown that, in some cases, redistribution of atoms not sputtered away by the ion-beam may be more important for nano-structure evolution than erosion itself \cite{norris-etal-NCOMM-2011}, which qualitatively altered the collective understanding of nano-structure formation. The ongoing, variegated development of theories comprising different combinations of prompt and gradual mechanisms, even for the low-energy irradiation of chemically-inert projectiles on elementally-pure semiconductors, speaks to the difficulty of forming a unified model. However, the viscous character of the amorphous layer has become well-established \cite{umbach-etal-PRL-2001,norris-etal-SREP-2017}.

One approach that has seen success as a tool in permitting the comparison of theory and experiment, and which has been used to great effect in early development of erosion-based theories as well as in the validation of the viscous flow model of noble gas irradiated silicon, is that of \textit{linear stability analysis} as in classic hydrodynamic stability work such as \cite{chandrasekhar-book-2013}. This is not to imply that linear stability analysis is applicable only to the hydrodynamic approach or gradual regime; they have also been done in the prompt regime \cite{bradley-harper-JVST-1988,bradley-PRB-2011b}. Within a linear stability analysis, a proposed model may be linearized about a steady-state configuration subject to small perturbations, leading to a system of differential equations which can be solved for a steady-state, and another system which can be solved for the perturbative correction. An appropriate Fourier-Laplace transform of the latter system leads to the linear dispersion relation, suggesting the growth-rate of variations in the surface depending on their wavenumber \cite{NorrisAziz_predictivemodel,norris-etal-SREP-2017}. In the case of models of ion-irradiated semiconductors, this is especially useful as it provides two predictions at the cost of one analysis. First, because wavenumbers for which the linear dispersion relation is negative imply the stability of the surface, while positivity implies instability, the most-unstable wavenumber is the one expected to be observed experimentally \cite{NorrisAziz_predictivemodel,munoz-garcia-etal-MSER-2014}. Second, the angle of irradiation naturally appears in the linear dispersion relation for these systems (as can be seen in \cite{castro-cuerno-ASS-2012,norris-PRB-2012-viscoelastic-normal,norris-PRB-2012-linear-viscous,moreno-barrado-etal-PRB-2015}), leading to a theoretical prediction for the transition from stability of the surface to perturbations to instability, the so-called critical angle or bifurcation angle, which is the minimal angle of the beam away from the vertical needed to begin to seen pattern formation. These two theoretical predictions can then be compared with any experimental data as validation (or repudiation) of theory. Indeed, the ability to correctly-predict experimentally observed wavelengths and the critical angle for transition is a minimal prerequisite for a successful theory. Such theoretical-experimental comparison was, in fact, instrumental in advancing the hypothesis of viscous relaxation as the primary regularizing mechanism in ion-irradiated thin films \cite{norris-etal-SREP-2017}.

The present work is the second part of a set of papers dedicated to the linear stability analysis of a stress-based model seeking to characterize the behavior of the amorphous layer. In the first part, we derived the long-wave linear dispersion relation for a highly-general model, sufficient to produce critical angle predictions, which considered two mechanisms, anisotropic plastic flow and isotropic swelling, with fully-arbitrary depth-dependence of each mechanism, which we derived from the projected down-beam direction in the form of Legendre polynomials. Without depth-dependence considerations, the anisotropic plastic flow model has previously led to good agreement between theory and experimental wavelength measurements in its full-spectrum form for 250eV Ar$^+ \to$ Si \cite{norris-PRB-2012-linear-viscous}, while isotropic swelling was recently proposed as an angle-independent stabilizing mechanism which could possibly explain the tendency for materials that undergo large amounts of ion-induced volumization to have higher critical angles \cite{Swenson_2018,evans-norris-JPCM-2022}. Comparison with a limited experimental data set also suggests that the anisotropic plastic flow model correctly identifies the form of the in-plane component of the steady-state stress tensor \cite{perkinsonthesis2017}, a successful theoretical prediction against other existing models of ion-induced flow \cite{castro-cuerno-ASS-2012,moreno-barrado-etal-PRB-2015}. However, in anticipation of the need to eventually apply these models to other experimental systems, possibly with very different physics, we are strongly motivated to generalization.

It was also shown that the assumption about the so-called ``interface relation" could lead to large changes in critical angle predictions, as much as 15 degrees even for very small swelling rates, between each of two interface relations \textit{which have recently been used for modeling by various theoretical groups}  \cite{castro-cuerno-ASS-2012,norris-PRB-2012-viscoelastic-normal,norris-PRB-2012-linear-viscous,Swenson_2018,evans-norris-JPCM-2022}. Because of the extreme variability in experimental predictions induced by these simple differences in modeling choices, we will seek to derive the correct interface relation directly from the physics of the ion-irradiated system, eliminating one degree of freedom from the analysis. We will then attempt to hypothesize as to the correct depth-dependence of each mechanism, conducting an analysis of the influence of the nuclear collision cascade in the amorphous bulk analogous to the analysis which originally gave rise to the prevailing surface-restricted model of erosion \cite{bradley-harper-JVST-1988,bradley-PRB-2011b}. This will serve as an adequate starting point for a physically-realistic, spatially-resolved model of the amorphous layer. 

The anisotropic plastic flow \cite{norris-PRB-2012-linear-viscous} and isotropic swelling models \cite{Swenson_2018,evans-norris-JPCM-2022} have been used, even until now, on a purely phenomenological basis, with a mechanism-agnostic approach taken to the underlying physics. The anisotropic plastic flow model, in particular, has been borrowed on a phenomenological basis from the electronic stopping regime, wherein a melt-cycle is thought to be the underlying physical mechanism \cite{van-dillen-etal-APL-2001-colloidal-ellipsoids,van-dillen-etal-APL-2003-colloidal-ellipsoids,van-dillen-etal-PRB-2005-viscoelastic-model,otani-etal-JAP-2006,wesch-wendler-book-2016}. However, such a melt-cycle does not occur within the nuclear stopping regime, our regime of concern, and the focus of many theoretical studies. Nonetheless, the associated stress tensor has led to good agreement between theory and experiment across many energy levels, and has seen widespread acceptance for these purposes \cite{trinkaus-ryazanov-PRL-1995-viscoelastic,trinkaus-NIMB-1998-viscoelastic,van-dillen-etal-APL-2001-colloidal-ellipsoids,van-dillen-etal-APL-2003-colloidal-ellipsoids,van-dillen-etal-PRB-2005-viscoelastic-model,otani-etal-JAP-2006,wesch-wendler-book-2016}. Isotropic swelling represents, phenomenologically, a generic volumizing mechanism which leads to isotropic stress production \cite{Swenson_2018,evans-norris-JPCM-2022}. Continuing with our broader goal of enhancing the physicality of our modeling, we will also show that our analysis and comparison with experimental and simulation results allow us to hypothesize as to the spatial distribution and physical origins of the anisotropic plastic flow and isotropic swelling mechanisms. We then compare parameters extracted from SRIM (``Stopping and Range of Ions in Matter", a simulation based on BCA, the ``binary collision approximation") \cite{ziegler-biersack-littmark-1985-SRIM} and experimental data for 30eV-100keV Ar$^+ \to$ Si \cite{madi-etal-2008-PRL,madi-etal-PRL-2011,madi-aziz-ASS-2012,moreno-barrado-etal-PRB-2015,hofsass-bobes-zhang-JAP-2016} and 400eV-2keV Xe$^+ \to$ Ge \cite{Teichmann2013} that appear to be consistent with our hypothesis. Understanding the physical origins and spatial variation of each mechanism will be helpful for any future attempts at parameter estimation via experimental (or MD) studies and for eventually predicting nano-structuring behavior from first principles at quantitative accuracy. 

\section{Model}
\begin{figure}[h!]
	\centering
	\includegraphics[totalheight=4.5cm]{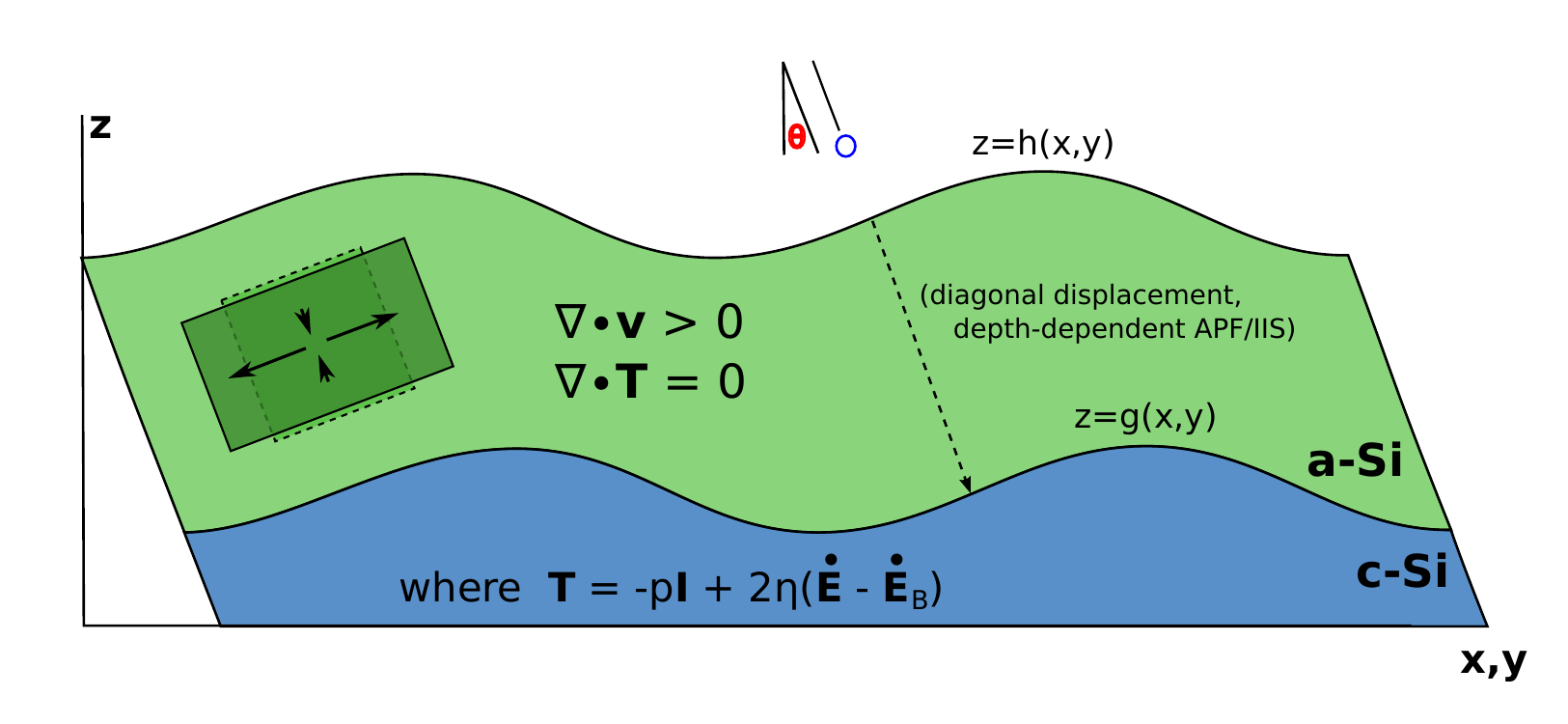}
	\caption{Schematic depicting ion bombardment at an incidence angle of $\theta$ and stress induced in the thin film by ion implantation.  Note that for off-normal incidence, the bottom boundary $z=g$ may not vertically align with the top boundary $z=h$. In this figure, we have made the simplifying assumption that all spatial variation occurs along the ``down-beam" direction.}
	\label{fig:schematic}
\end{figure}

\subsection{Preliminaries}
We will briefly review the relevant physical mechanisms here. For a more complete discussion, we refer the reader to our previous work, the first part of the present set.

\paragraph{Anisotropic plastic flow (APF).} The proposed source of anisotropy and angle-dependent destabilization of the surface is, phenomenologically, that of anisotropic plastic flow, associated with the stress tensor in Equation (\ref{APFtensor}). In the electronic stopping regime, the form of this stress tensor is associated with the tendency of the region immediately around the ion-track to rapidly heat, expand, and cool, ``freezing-in" a deformation whose effect is to compress a parcel of matter in the downbeam direction and to expand it in the crossbeam direction \cite{trinkaus-ryazanov-PRL-1995-viscoelastic,trinkaus-NIMB-1998-viscoelastic,van-dillen-etal-APL-2001-colloidal-ellipsoids,van-dillen-etal-APL-2003-colloidal-ellipsoids,van-dillen-etal-PRB-2005-viscoelastic-model,otani-etal-JAP-2006,wesch-wendler-book-2016}. This leads to its occasional description as ``pancake-strain" \cite{NorrisAziz_predictivemodel,perkinsonthesis2017}. A morphological instability results as these anisotropically strained components occur beyond the critical angle of the beam. For more discussion, see \cite{norris-PRB-2012-linear-viscous,wesch-wendler-book-2016}. However, in the nuclear stopping regime, there is not enough energy allocated to inelastic collisions to produce the thermal spike and melt-cycle that typically motivate the form of the stress tensor \cite{van-dillen-etal-APL-2001-colloidal-ellipsoids,van-dillen-etal-APL-2003-colloidal-ellipsoids,van-dillen-etal-PRB-2005-viscoelastic-model,norris-PRB-2012-linear-viscous,wesch-wendler-book-2016}. The form of the stress tensor, borrowed directly from the electronic stopping regime, has shown ``unreasonably good" agreement with experiment across numerous applications \cite{van-dillen-etal-APL-2001-colloidal-ellipsoids,van-dillen-etal-APL-2003-colloidal-ellipsoids,van-dillen-etal-PRB-2005-viscoelastic-model,otani-etal-JAP-2006,george-etal-JAP-2010,norris-PRB-2012-linear-viscous,}, and has become fairly widespread in its use in a phenomenological basis. Nonetheless, the underlying physics is currently a point of speculation. It appears that \textit{something} analogous to the melt-cycle of the electronic stopping regime occurs in the nuclear stopping regime, but its nature is, as yet, undetermined. 

\paragraph{Ion-induced swelling (IIS).} Although from an experimental standpoint, swelling due to radiation damage is well-documented and well-studied at high energies (see \cite{Holland1983,Wang1989,Jafri1989,McHargue1993,Tamulevicius1996,Giri2001,Boettger2013} and many others), it was only recently studied phenomenologically in the low-energy ion-irradiation context \cite{Swenson_2018,evans-norris-JPCM-2022}. This model supposes that as a parcel of matter moves throughout the amorphous film, it accumulates damage and begins to volumize. The longer a parcel of matter remains within the film, the more volumization occurs. This isotropic swelling induces a stabilizing effect, allowing the upper interface to more effectively ``recover" from perturbations, as the volumization appears at the upper interface as lateral expansion. For long-wave perturbations, this helps to ``fill in" the valleys. For more discussion, see \cite{Swenson_2018,evans-norris-JPCM-2022}, where the swelling rate is assumed constant throughout the film. 

\paragraph{Objectives of the present work.} In iur previous work, isotropic swelling, like anisotropic plastic flow, was allowed to vary spatially in a vanishingly-thin ion-track. Due to the amorphization of the underlying crystalline substrate, this ion-track would characterize the relationship between the free upper interface and the lower amorphous-crystalline interface. However, this simple idealization appears to vastly underpredict $\theta_c$, the minimal irradiation angle for the transition from flat surfaces to patterned surfaces. For contrast, we considered an alternative idealization, that of spatial variation along the laboratory z axis and the amorphous-crystalline interface being a vertical translation of the free interface regardless of beam angle. This idealization overpredicts $\theta_c$. Curiously, both idealizations exist within the literature and have been used for comparison with experimental and MD simulation data. We hypothesize that the widespread use of these, and other, idealizations may be a significant source of disagreement between experimental work and the models advanced by different groups. In the present work, we will significantly improve the physical realism of our modeling: (1) deriving highly-accurate closed-form approximations for the shape of the lower interface in terms of the upper, which are then verified against quasi-empirical film thickness data; and (2) carrying out a spatially-resolved analysis in terms of power deposition within the amorphous bulk approximated by a bivariate Gaussian ellipsoid in downbeam-crossbeam coordinates, as has been done elsewhere \cite{sigmund-PR-1969,sigmund-JMS-1973,bradley-harper-JVST-1988}. Having done so, we will be in an excellent position to incorporate data obtained from simulation tools, such as SRIM \cite{ziegler-biersack-littmark-1985-SRIM}, directly into any future linear stability analysis. We expect that these enhancements will promote the alignment of theory and experiment, especially for critical angle and wavelength selection.

\subsection{Governing equations}
As in our previous work, we adopt the following notation: $\rho$ denotes density, $\vec{v}$ denotes the velocity field, with $u$ along the x-axis and $w$ along the z-axis. $p$ denotes pressure, $\theta$ denotes the nominal angle of the beam to the surface in laboratory coordinates (see Figure \ref{fig:schematic}). $\Delta$ tracks the local ``volumization" field. We have used $\Delta$ as a mnemonic for ``change" suggested by $\Delta V$, a typical notation for ``change in volume". However, we already have a quantity $V$ here: it denotes the steady-state translation of the film. $\rho^*$ denotes the original crystalline density of the substrate, prior to amorphization. $V_{I,h}$ is the interfacial velocity; $\hat{n}$ the outward-normal unit vector to the surface, and $\hat{t}$ a unit tangent vector to the surface. $\hat{k}$ is the unit direction vector along the laboratory $z$ axis. $\gamma$ is a surface energy constant. $\kappa$ denotes local curvature. In the bulk, we have the description of mass and momentum conservation
\begin{equation}
\begin{gathered}
	\frac{\partial \rho}{\partial t} + \nabla \cdot (\rho \vec{v}) = 0 \\
	\nabla \cdot \textbf{T} = 0,
\end{gathered}
\end{equation}
which is nearly a description of classical Stokes flow, except for modification of the momentum equations by
\begin{equation} \label{APFtensor}
	\begin{gathered}
		\textbf{T} = - p\textbf{I} + 2\eta\{\dot{\textbf{E}} - \dot{\textbf{E}}_b\} \\
		\dot{\textbf{E}} = \frac{1}{2}\left( \nabla \vec{v} + \nabla \vec{v}^T \right) \\
		\dot{\textbf{E}}_b = fA\tau(z;...) \textbf{D}(\theta) \\
		\textbf{D}(\theta) = 
		\begin{bmatrix} 
			\frac{3}{2}\cos(2\theta) - \frac{1}{2} & 0 & \frac{3}{2}\sin(2\theta) \\
			0 & 1 & 0 \\
			\frac{3}{2}\sin(2\theta) & 0 & -\frac{3}{2}\cos(2\theta) - \frac{1}{2} \\
		\end{bmatrix},
	\end{gathered}
\end{equation}
which incorporates, on a phenomenological basis, an angle-dependent stress-free strain rate. In order to track volumization, again on a mechanism-agnostic basis, we take
\begin{equation}
	\begin{gathered}
		\frac{\partial \Delta}{\partial t} + \vec{v}\cdot \nabla(\Delta) = \alpha(z;...) \\
		\rho = \frac{\rho^*}{1+\Delta}.
	\end{gathered}
\end{equation}

\noindent At the free upper interface, $z=h$, we have
\begin{equation}
	\begin{gathered}
		v_{I,h} = \vec{v}\cdot \hat{n} - V\frac{\rho^*}{\rho} \\
		\textbf{[T]} \cdot \hat{n} = -\gamma \kappa \hat{n},
	\end{gathered}
\end{equation}

The first equation is a modified kinematic condition which takes into account the removal of material due to erosion. At the amorphous-crystalline interface, $z=g$, we have
\begin{equation}
	\begin{gathered}
		\Delta = 0 \\
		\vec{v}\cdot \hat{t} = 0, \\
		\vec{v}\cdot \hat{n} = 0,
	\end{gathered}
\end{equation}
which are simply the no-slip and no-penetration conditions at the amorphous-crystalline boundary and the statement that newly-amorphized Si has not had time to ``volumize" upon entering the bulk. Because the film is eroding in the downbeam direction with time, we convert to a moving frame via
\begin{equation}
	\begin{gathered}
		h \to h-Vt \\
		g \to g-Vt \\
		v_{I,h} \to v_{I,h} - V(\hat{k}\cdot\hat{n}) \\
		z \to z-Vt \\
		\vec{v} \to \vec{v} - V\hat{k}.
	\end{gathered}
\end{equation}

\section{Analysis}
Our earlier work has obtained the long-wave linear dispersion relation,
\begin{equation}
	\sigma \approx 0 + k\left(\sigma_{10} + \hat{\alpha}\sigma_{11} \right) + k^2\left(\sigma_{20} + \hat{\alpha}\sigma_{21} \right),
\end{equation}
with
\begin{equation}
	\begin{gathered}
		\sigma_{10} = -\frac{2fAiD_{13}}{\tilde{h}_1}\Bigg[\int_0^{h_0}\int_0^{z_1}\tau_{\epsilon}(z_2)dz_2dz_1 - \tau_0(0)\tilde{g}_1h_0 + \int_0^{h_0}\tau_0(z)dz\Bigg];
	\end{gathered}
\end{equation}


\begin{equation}
	\begin{gathered}
		\sigma_{11} = \frac{1}{V\tilde{h}_1}\int_0^{h_0}\Bigg[-2fAiD_{13}\alpha_{10}(z)\Big[\int_0^{z}\int_0^{z_1}\tau_{\epsilon}(z_2)dz_2dz_1 - \tau_0(0)\tilde{g}_1z\Big] \\
		+ \sigma_{10}\Big[\int_0^{z}\alpha_{1\epsilon}(z_1)dz_1 - \alpha_{10}(0)\tilde{g}_1\Big]\Bigg]dz;
	\end{gathered}
\end{equation}


\begin{equation}
	\begin{gathered}
		\sigma_{20} = \frac{2fA(D_{11}-D_{33})}{\tilde{h}_1}\int_0^{h_0}\Bigg[\int_0^{z}\int_0^{z_1}\tau_{\epsilon}(z_2)dz_2dz_1 - z\Big[\tau_0(h_0) + \int_0^{h_0}\tau_{\epsilon}(z)dz\Big] \Bigg]dz;
	\end{gathered}
\end{equation}
and

\begin{equation}
	\begin{gathered}
		\sigma_{21} = -\frac{1}{\tilde{h}_1}\int_0^{h_0}\Bigg[\frac{\sigma_{10}}{V^2}\int_0^{z}\Big[ \Big(\int_0^{z_1}\alpha_{1\epsilon}(z_2)dz_2 - \alpha_{10}(0)\tilde{g}_1\Big)\Big(\sigma_{10} + 2fAiD_{13}\int_0^{z_1}\tau_0(z_2)dz_2\Big) \\
		- 2fAiD_{13}\alpha_{10}(z_1)\Big(\int_0^{z_1}\int_0^{z_2}\tau_{\epsilon}(z_3)dz_{2} - \tau_0(0)\tilde{g}_1z_1\Big)\Big]dz_1 \\
		-
		\frac{\sigma_{20}}{V}\Big[\int_0^z\alpha_{1\epsilon}(z_1)dz_1 - \alpha_{10}(0)\tilde{g}_1\Big] \\
		- \Big[\int_0^z\int_0^{z_1}\alpha_{1\epsilon}(z_2)dz_2dz_1 + z\Big(\alpha_{10}(0)\tilde{g}_1 - 2\alpha_{10}(h_0) - 2\int_0^{h_0}\alpha_{1\epsilon}(z)dz\Big)\Big] \\
		- \frac{2fA\alpha_{10}(z)(D_{11}-D_{33})}{V}\Big[\int_0^z\Big(\int_0^{z_1}\int_0^{z_2}\tau_{1\epsilon}(z_3)dz_3dz_2 - z_1(\tau_0(h_0) + \int_0^{h_0}\tau_{1\epsilon}(z)dz)\Big)dz_1\Big]\Bigg]dz
	\end{gathered}
\end{equation}
It is useful to notice that in the limit of small cross-terms, these expressions can be simplified greatly; in particular,
\begin{equation} \label{dispnreln1}
	\begin{gathered}
		\sigma_{10} = -\frac{2fAiD_{13}}{\tilde{h}_1}\Bigg[\int_0^{h_0}\int_0^{z_1}\tau_{\epsilon}(z_2)dz_2dz_1 - \tau_0(0)\tilde{g}_1h_0 + \int_0^{h_0}\tau_0(z)dz\Bigg];
	\end{gathered}
\end{equation}

\begin{equation}
	\sigma_{11} = 0;
\end{equation}

\begin{equation}
	\sigma_{20} = \frac{2fA(D_{11}-D_{33})}{\tilde{h}_1}\int_0^{h_0}\Bigg[\int_0^{z}\int_0^{z_1}\tau_{\epsilon}(z_2)dz_2dz_1 - z\Big[\tau_0(h_0) + \int_0^{h_0}\tau_{\epsilon}(z)dz\Big]\Bigg]dz
\end{equation}
and

\begin{equation} \label{dispnreln2}
	\begin{gathered}
		\sigma_{21} = \frac{1}{\tilde{h}_1}\int_0^{h_0}\Bigg[\int_0^z\int_0^{z_1}\alpha_{1\epsilon}(z_2)dz_2dz_1 +z\Big[\alpha_{10}(0)\tilde{g}_1 - 2\alpha_{10}(h_0) - 2\int_0^{h_0}\alpha_{1\epsilon}(z)dz\Big]     \Bigg],
	\end{gathered}
\end{equation}
where the $0$ subscript for $\alpha_1$ and $\tau$ denotes the steady-state description of these intensity fields and the $\epsilon$ subscript denotes the perturbative correction to the steady-state intensity field for a thin film under a small perturbation at the upper interface. Based on our previous work, it appears that a pervasive lack of physical realism has been a major factor in the ongoing difficulties in reconciling theory and experiment. We will therefore connect the above functional forms to more physically-realistic descriptions of (1) the bulk description of spatial variation of APF and IIS and (2) the interface relation, which characterizes the displacements between the free interface and the amorphous-crystalline interface, both of which may be immediately incorporated into the functionals.

\subsection{Interface relation from BCA}
\begin{figure}[h!]
	\centering
	\includegraphics[totalheight=4.5cm]{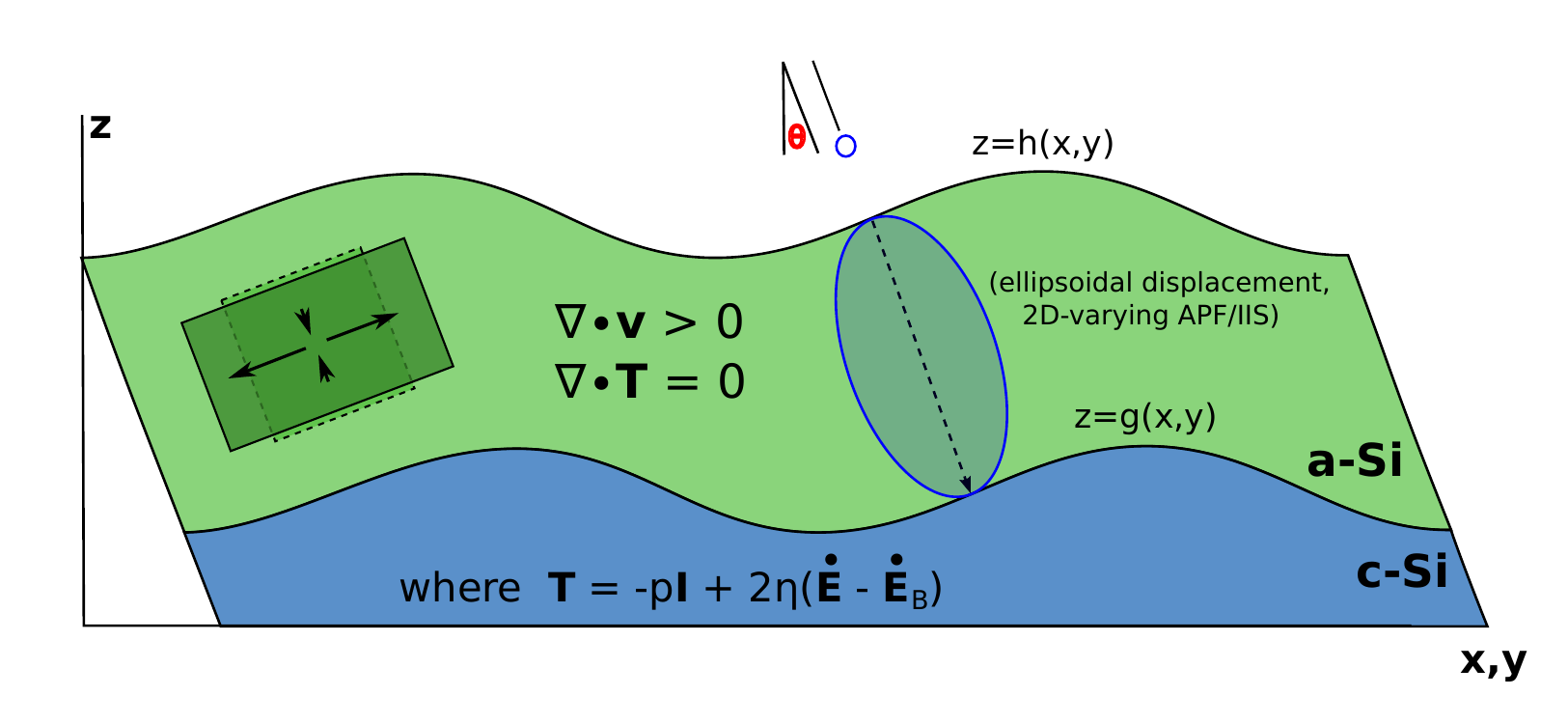}
	\caption{Schematic depicting ion bombardment at an incidence angle of $\theta$ and stress induced in the thin film by ion implantation.  Note that for off-normal incidence, the bottom boundary $z=g$ may not vertically align with the top boundary $z=h$. Based on the Gaussian ellipsoid associated with power deposition averaged in time across multiple ion-impacts, we will consider (1) the spatial variation of stresses associated with APF and IIS and (2) the location of the lower interface. Notice that if the Gaussian ellipsoid rotates with the beam angle, we can obtain an interface that is neither the vertical translation of \cite{norris-PRB-2012-linear-viscous,norris-PRB-2012-viscoelastic-normal} nor the pure diagonal translation of \cite{moreno-barrado-etal-PRB-2015,Swenson_2018,evans-norris-JPCM-2022} by determining the lowest point on the laboratory $z$ axis amorphized by the ellipsoid (or, more precisely, the location for which its tangent plane is parallel to the tangent plane at the surface where the ion entered).}
	\label{fig:schematic2}
\end{figure}

From Part 1 of the present series, we have seen that different assumptions about the interface relation may lead to significant differences in predictions for $\theta_c$. Seeking to eliminate a degree of freedom from our analysis, we now derive a closed-form expression relating the interface relation directly to details that can be obtained readily from BCA simulation, such as SRIM. From SRIM, it is straightforward to obtain the parameters $a$, $\alpha$ and $\beta$, which characterize the Gaussian ellipsoidal approximation of the deposited ions: $a$ is the mean penetration depth, $\alpha$ is the downbeam standard deviation (``straggle") and $\beta$ is the crossbeam standard deviation (``straggle"). At low energies, this will serve as an approximation of power deposition; our results will confirm that the approximation is sufficient for the purposes of the present work.

\paragraph{Derivation.} Without loss of generality, we consider the influence of the collision cascade generated by ions entering the film through a single patch of surface at $(0,0)$ with local incidence angle $\phi$. In three dimensions, we can determine, in general, all tangent planes to some level curve of the Gaussian ellipsoid where there has been enough power deposition to amorphize the film (see Figure \ref{fig:schematic2}). We then seek such a tangent plane parallel to the tangent plane at the patch of surface, and we identify the displacement along the $z$ axis between these two planes as a predicted local film thickness. The change in the $(x,y)$ plane between the patch of surface and the Gaussian ellipsoid where its tangent plane rests may be identified with the local lateral shift in the interfaces. We will then seek a general expression for the local film thickness and the local horizontal shift. Here, the two-dimensional case if sufficient, which simplifies the calculations. With the classical approximation of energy deposition by a bivariate Gaussian in crossbeam-downbeam coordinates $(\tilde{x},\tilde{z})$ \cite{sigmund-PR-1969,sigmund-JMS-1973,bradley-harper-JVST-1988,bradley-PRB-2011b}, we define
\begin{equation}
	E_D(\hat{x},\hat{z}) = E_0\exp\left(-\frac{(\tilde{z}-a)^2}{2\alpha^2} - \frac{\tilde{x}^2}{2\beta^2}\right),
\end{equation}
where $E_D(\hat{x},\hat{z})$ denotes the energy deposition at a location in crossbeam-downbeam coordinates. We transform these coordinates into Cartesian via
\begin{equation}
	\begin{gathered}
		\tilde{x} = x\cos(\phi) - z\sin(\phi) \\
		\tilde{z} = z\cos(\phi) + x\sin(\phi).
	\end{gathered}
\end{equation}
However, because $\tilde{z}$ was originally downbeam, hence increasing $\tilde{z}$ reflects a ``downward" change in position against ``typical" (x,z) coordinates (which we seek to recover), we now take
\begin{equation}
	z \to -z.
\end{equation}
We compute level curves such that
\begin{equation}
	E_D = E_A,
\end{equation}
that is, the deposited energy per unit is equal to the energy per unit required to induce amorphization. Hence
\begin{equation}
	f(x,z) = \frac{1}{2}\left(\frac{(a+z\cos(\phi)-x\sin(\phi))^2}{\alpha^2} + \frac{(x\cos(\phi)+z\cos(\phi))^2}{\beta^2}   \right) - \ln\left(\frac{E_0}{E_A}\right),
\end{equation}
and we seek the locations on this level curve such that
\begin{equation}
	-\frac{\nabla f}{||\nabla f||} = <0,1>, 
\end{equation}
so that the normal vector to the level curve points ``straight down". This is essentially a simplification of the condition that $-\frac{\nabla f}{||\nabla f||} = \hat{n}$, where $\hat{n}$ is the outward-normal unit vector at the location on the surface at which the ion entered the film. In the case of a long-wave linear stability analysis conducted about a surface that is ``nearly flat" in the reference frame, the simplification will turn out to be sufficient. We then focus on the x component for ease of solution, leading to
\begin{equation}
	x = \frac{a\beta^2\sin(\phi) - z\alpha^2\cos(\phi)\sin(\phi) + z\beta^2\cos(\phi)\sin(\phi)  }{\alpha^2\cos^2(\phi) + \beta^2\sin^2(\phi)};
\end{equation}
substitution into the equation for the level curve yields an equation purely in $z$, from which values of $z$ can be obtained. This leads to
\begin{equation}
	x = a\sin(\phi) \pm 2(\alpha^2-\beta^2)\sin(\phi)\cos(\phi)\sqrt{\left(\frac{\ln(\frac{E_0}{E_A})}{\alpha^2+\beta^2 + (\alpha^2-\beta^2)\cos(2\phi)} \right)  }
\end{equation}
and
\begin{equation}
	z = -a\cos(\phi) \mp 2\left(\left[\alpha^2\cos^2(\phi) + \beta^2\sin^2(\phi)\right]\sqrt{\left(\frac{\ln(\frac{E_0}{E_A})}{\alpha^2+\beta^2 + (\alpha^2-\beta^2)\cos(2\phi)} \right)} \right) 
\end{equation}
and we must choose the pair that gives the physical solution we care about; in particular, the ``+" solution for x and the ``-" solution for z. Because we took the origin (the entry point of the ion) as (0,0), this (x,z) pair signifies the displacement from the origin (i.e., surface) of the corresponding point on the lower interface. Identifying that interface with $g(x,t)$, we have

\begin{equation}
	x_0 = a\sin(\phi) + 2(\alpha^2-\beta^2)\sin(\phi)\cos(\phi)\sqrt{\left(\frac{\ln(\frac{E_0}{E_A})}{\alpha^2+\beta^2 + (\alpha^2-\beta^2)\cos(2\phi)} \right)  }
\end{equation}
and
\begin{equation}
	h_0 = a\cos(\phi) + 2\left(\left[\alpha^2\cos^2(\phi) + \beta^2\sin^2(\phi)\right]\sqrt{\left(\frac{\ln(\frac{E_0}{E_A})}{\alpha^2+\beta^2 + (\alpha^2-\beta^2)\cos(2\phi)} \right)} \right) 
\end{equation}
such that
\begin{equation}
	g(x,t) = h(x-x_0(\phi),t) - h_0(\phi).
\end{equation}
It can be shown that these forms reduce to
\begin{equation}
	\begin{gathered}
		h_0 = a\cos(\phi) + 2\left(\sqrt{\alpha^2\cos^2(\phi) + \beta^2\sin^2(\phi)}\sqrt{\frac{\ln(\frac{E_0}{E_A})}{2}}   \right) \\
		x_0 = a\sin(\phi) + 2\left(\frac{(\alpha^2-\beta^2)\sin(\phi)\cos(\phi)}{\sqrt{\alpha^2\cos^2(\phi) + \beta^2\sin^2(\phi)}}\sqrt{\frac{\ln(\frac{E_0}{E_A})}{2}}   \right).
	\end{gathered}
\end{equation}
If we then assume that the amorphization threshold energy $E_A$ is well-approximated by two standard deviations of energy deposition in the cross-beam direction away from the center of the collision cascade, as has been discussed elsewhere \cite{hoffsass-etal-APA-2012,norris-etal-SREP-2017}, we have (in downbeam-crossbeam) $\tilde{z}=a$, $\tilde{x} = 2\beta$, hence, from the energy deposition function, 
\begin{equation}
	E_A = E_0\exp\left(-2\right) \to \ln(\frac{E_0}{E_A}) = 2,
\end{equation}
and the above further reduce to
\begin{equation}
	\begin{gathered}
		h_0 = a\cos(\phi) + 2\left(\sqrt{\alpha^2\cos^2(\phi) + \beta^2\sin^2(\phi)}\right) \\
		x_0 = a\sin(\phi) + 2\left(\frac{(\alpha^2-\beta^2)\sin(\phi)\cos(\phi)}{\sqrt{\alpha^2\cos^2(\phi) + \beta^2\sin^2(\phi)}}\right),
	\end{gathered}
\end{equation}

\paragraph{Error estimate; compatibility with linear stability analysis.}
While this is true for an arbitrary patch of irradiated (but otherwise flat) surface, we wish to consider the case where an entire, possibly non-flat surface is uniformly irradiated. It is natural and simple to consider the small-slopes expansion of the above local angle. That is,
\begin{equation}
	\phi = \theta - h_x,
\end{equation}
which leads to
\begin{equation}
	\begin{gathered}
		h_0 = a\cos(\theta) + 2\left(\sqrt{\alpha^2\cos^2(\theta) + \beta^2\sin^2(\theta)}\right) +
		h_x\left(a\sin(\theta) + \frac{(\alpha^2-\beta^2)\sin(2\theta)}{\sqrt{\alpha^2\cos^2(\theta) + \beta^2\sin^2(\theta)}} \right) + O(h_x^2)
	\end{gathered}
\end{equation}
and
\begin{equation}
	\begin{gathered}
		x_0 = a\sin(\theta) + 2\left(\frac{(\alpha^2-\beta^2)\sin(\theta)\cos(\theta)}{\sqrt{\alpha^2\cos^2(\theta) + \beta^2\sin^2(\theta)}}\right) \\
		- h_x\left(a\cos(\theta) + \frac{(\alpha^2-\beta^2)\left(4(\alpha^2+\beta^2)\cos(2\theta) + (\alpha^2-\beta^2)(3+\cos(4\theta)) \right)}{ (\alpha^2\cos^2(\theta) + \beta^2\sin^2(\theta))^{3/2} }           \right) + O(h_x^2).
	\end{gathered}
\end{equation}
Usefully, these quantities are constants to leading order, and with the small-perturbation expansion,
\begin{equation}
	h \to h_0 + \epsilon h_1(x,t), g \to 0 + \epsilon g_1(x,t).
\end{equation}
Identifying
\begin{equation}
	g_1(x,t) = h_1(x-x_0,t) - h_0,
\end{equation}
it soon becomes apparent that the influence of the term $\propto h_x$ in each of $x_0$ and $h_0$ induces another $O(\epsilon)$ correction, itself multiplied by $\epsilon$ due to the linearization. Hence for the purposes of the present work, conducted to $O(\epsilon)$, we may safely neglect the small-slope correction and simply take
\begin{equation}
	\begin{gathered}
		x_0 = a\sin(\theta) + 2\left(\frac{(\alpha^2-\beta^2)\sin(\theta)\cos(\theta)}{\sqrt{\alpha^2\cos^2(\theta) + \beta^2\sin^2(\theta)}}\right); \\
		h_0 = a\cos(\theta) + 2\left(\sqrt{\alpha^2\cos^2(\theta) + \beta^2\sin^2(\theta)}\right)		
	\end{gathered}
\label{thicknesseqn}
\end{equation}
throughout our analysis, as the next correction for small perturbations occurs at $O(\epsilon^2)$.

\paragraph{Some interesting limits.}
We note that the above expressions completely reproduce the previously-studied interface relations, which are shown to be special cases of the above. In particular, the ``vanishing cross-beam width" limit,
\begin{equation}
	\begin{gathered}
		\lim_{\beta \to 0}h_0(\theta) = (a+2\alpha)\cos(\theta) \\
		\lim_{\beta \to 0}x_0(\theta) = (a+2\alpha)\sin(\theta), \\
	\end{gathered}
\end{equation}
is the ``pencil-cascade" relation used in \cite{moreno-barrado-etal-PRB-2015,Swenson_2018,evans-norris-JPCM-2022}. A particular ``no angle-dependence" limit,
\begin{equation}
	\begin{gathered}
		\lim_{\theta \to 0}h_0(\theta) = (a+2\alpha) \\
		\lim_{\theta \to 0}x_0(\theta) = 0, \\
	\end{gathered}
\end{equation}
is the ``vertical-translation" case used in \cite{norris-PRB-2012-linear-viscous,norris-PRB-2012-viscoelastic-normal}. The simple case of perfectly spherical collision cascades, $\alpha = \beta$, is
\begin{equation}
	\begin{gathered}
		\lim_{\beta \to \alpha}h_0(\theta) = a\cos(\theta) + 2\alpha \\
		\lim_{\beta \to \alpha}x_0(\theta) = a\sin(\theta), \\
	\end{gathered}
\end{equation}
which is qualitatively between the vertical translation case and the fully general collision cascade relation. The spherical approximation has been previously considered at least by \cite{carter-vishnyakov-PRB-1996,cuerno-barabasi-PRL-1995} as a simplifying assumption broadly appropriate for low-energy noble gas ion irradiation of Si. We note that the case of a flat lower interface of \cite{castro-cuerno-ASS-2012,munoz-garcia-etal-PRB-2019} has \textit{no} associated limit within this geometric derivation. Our derivation has implicitly assumed fast or instantaneous amorphization (i.e., fast or instantaneous power deposition beyond the amorphization threshold \cite{priolo-etal-ASS-1989,belyakov-titov-REDS-1996,titov-kucheyev-NIMB-2000,titov-etal-NIMB-2003,titov-etal-AVPS-2003,titov-etal-TSF-2006,titov-etal-PRB-2006,oliviero-etal-JAP-2006,wesch-wendler-book-2016,okulich-etal-FTTMM-2020}; in the most general case, we expect that the interfaces would be obtained from a system of partial differential equations tracking power deposition and defect accumulation, perhaps generalizations of the point-models of \cite{belyakov-titov-REDS-1996,titov-kucheyev-NIMB-2000,titov-etal-NIMB-2003,titov-etal-AVPS-2003,titov-etal-TSF-2006,titov-etal-PRB-2006}. Strictly speaking, such an approximation may not hold in the case of very low fluxes, as the time in between successive collision events leading to power deposition at a given $(x,z)$ location could be greater than the time required for the damage to anneal away. However, under such circumstances, we would not expect there to be an amorphous layer to speak of--- and then our model (as with all other hydrodynamic-type models) would be \textit{broadly} inapplicable. It is also possible, in principle, that our derivation of the shape of the amorphous-crystalline interface is no longer valid for systems that are poorly-approximated by Gaussian ellipsoidal power deposition. Indeed, as has been discussed in \cite{hobler-etal-PRB-2016}, power deposition within the irradiated film may not perfectly follow a Gaussian distribution near grazing incidence. There may also be significant errors in the Gaussian approximation for films with large variations in their density \cite{hossain-etal-JAP-2012}. Nonetheless, the present approach of ``determining the level curve of the power-deposition function necessary to induce amorphization and then identifying the interfacial displacements" should be preserved as long as amorphization is sufficiently rapid, regardless of the \textit{exact} form of power deposition.

\paragraph{Implications for parameter estimation.} 
In \cite{perkinsonthesis2017}, estimates of stress were obtained using Stoney's equation,
\begin{equation}
	\sigma_{avg} = \frac{\Delta K M_s h_s^2}{6h_f}.
\end{equation}
Here, $\sigma_{avg}$ denotes the averaged in-plane stress within the film, $\Delta K$ denotes the change in curvature, $M_s$ is the biaxial modulus of the material, $h_s$ is the thickness of the substrate, and $h_f$ is the thickness of the amorphous layer. In the original stress measurements of \cite{perkinsonthesis2017}, the film thickness was assumed to be 3nm throughout all irradiation angles (i.e., the ``vertical interfaces" assumption described earlier). Because we now have a physically-accurate (as will be shown) expression for the film-thickness, we are able to ``back-step" the value of $\Delta K$ and re-fit the value of $\sigma_{avg}$ to a modified value which includes the angular-dependence of the film thickness. As we will also show, this correction leads to a significant change in the values of $fA_I\eta$ and $fA_D\eta$. This highlights the importance of carefully-implemented physical detail.

\subsection{Ion-induced stress distributions from BCA}
As earlier, we take downbeam mean $a$, downbeam standard deviation $\alpha$ and crossbeam standard deviation $\beta$, with $(\hat{x},\hat{z})$ as crossbeam-downbeam coordinates respectively. Then we have
\begin{equation}
	E_D(\tilde{x},\tilde{y},\tilde{z}) = \frac{1}{(2\pi)^{3/2}\alpha\beta^2}\exp\left(-\frac{(\tilde{z}-a)^2}{2\alpha^2} - \frac{\tilde{x}^2+\tilde{y}^2}{2\beta^2} \right).
\end{equation}
To restrict attention to the $\tilde{x},\tilde{z}$ plane before transformation to $(x,z)$, we integrate out $\tilde{y}$ and obtain
\begin{equation}
	E_D(\tilde{x},\tilde{z}) = \frac{1}{2\pi\alpha\beta}\exp\left(-\frac{(\tilde{z}-a)^2}{2\alpha^2} - \frac{\tilde{x}^2}{2\beta^2} \right),
\end{equation}
and we may change from crossbeam-downbeam coordinates to typical Cartesian, while simultaneously choosing $(X,h(X))$ as the origin of the incoming ion, via the transformation
\begin{equation}
	\begin{gathered}
		\tilde{x} \to (x-X)\cos(\theta) + (z-h(X))\sin(\theta) \\
		\tilde{z} \to (x-X)\sin(\theta) - (z-h(X))\cos(\theta).
	\end{gathered}
\end{equation}
This prepares us to consider the linearization of the surface in laboratory (x,z) coordinates. We then have, in (x,z),
\begin{equation}
	\begin{gathered}
		E_D(x,z) =  \frac{1}{2\pi\alpha \beta}\exp\left(-\frac{[(x-X)\sin(\theta) - (z-h(X))\cos(\theta)-a]^2}{2\alpha^2} - \frac{[(x-X)\cos(\theta) + (z-h(X))\sin(\theta)]^2}{2\beta^2} \right),
	\end{gathered}
\end{equation}
an expression of instantaneous energy deposition at the point $(x,z)$ due to ions incoming from $(X,h(X))$ on the surface. But the calculation of mean power deposition throughout the bulk due to a possibly variable surface must be flux-weighted by local surface slope against the nominal ion beam angle. This is because of the geometric effect of slopes off-normal against the ion-beam direction effectively creating a larger effective area over which the same number of ions are uniformly allocated. Then in calculating mean deposited power over time, we must evaluate
\begin{equation}
	P(x,z) = \int_{-\infty}^{\infty}\left(\cos(\theta) + h_X\sin(\theta)\right)\times E_D(x,z) dX.
\end{equation}
We note that our discussion up to this point closely follows that of \cite{bradley-harper-JVST-1988,bradley-PRB-2011b}. Clearly, the above integral cannot be evaluated in closed form for arbitrary $h(X)$, and we are specifically interested in closed-form solutions. Instead, we linearize in a near-flat $h(X)$ and then carry out the integration for the linearized integrand; because we intended to seek a linear stability result from the beginning, we restore analytical tractability at no penalty to value the analysis. We take
\begin{equation}
	h(X) = h_0  + \epsilon h_1(X), 
\end{equation}
which leads to the expression for the integrand of the above,
\begin{equation}
	\begin{gathered}
		I = \frac{1}{2\pi\alpha\beta}\cos(\theta)I_0 + \\ \epsilon\frac{\cos(\theta)}{2\pi\alpha^3\beta^3}\Big[h_1(X)I_0\Big((z-h_0)(\alpha^2\sin^2(\theta)+\beta^2\cos^2(\theta)) + \cos(\theta)a\beta^2 +  \cos(\theta)\sin(\theta)(x-X)(\alpha^2-\beta^2)\Big) \\ + \alpha^2\beta^2\tan(\theta)h_1'(X)\Big]  \\
		+ O(\epsilon^2),
	\end{gathered}
\end{equation}
where
\begin{equation}
	I_0 = \exp\left(-\frac{[(x-X)\sin(\theta) - (z-h_0)\cos(\theta)-a]^2}{2\alpha^2} - \frac{[(x-X)\cos(\theta) + (z-h_0)\sin(\theta)]^2}{2\beta^2} \right)
\end{equation}
Next we apply the linear stability ansatz to the above in spatial variable $X$, retaining $x$ as a fixed constant for the moment, being the $x$ coordinate at which we wish to determine mean power deposition. Hence
\begin{equation}
	h_1(X) \to \tilde{h}_1\exp(ikX+\sigma t).
\end{equation}
We now wish to integrate the linearized quantity $I$ from $X=-\infty$ to $X=+\infty$ in order to account for the flux-diluted power deposition at a fixed $(x,z)$ due to ions entering through \textit{all} points $(X,h(X))$. We make a change of variables $p := x-X$, so $dX = -dp$. The linearization combined with this change of variables allows us to consider
\begin{equation}
	\exp(ikX) = \exp(ikx+\sigma t)\exp(-ikp);
\end{equation}
this is useful because we expect the cancellation of a factor of $\exp(ikx+\sigma t)$ so that we can solve the linearized differential equations in the $z$ coordinate alone, as is typical in such calculations. Now this leads to
\begin{equation}
	\begin{gathered}
		P(x,z) = \frac{\cos(\theta)}{2\pi\alpha\beta}\int_{-\infty}^{\infty}I_0dp \\
		+ \epsilon \tilde{h}_1\exp(ikx + \sigma t) \frac{\cos(\theta)}{2\pi \alpha^3 \beta^3}  \left(c_1\int_{-\infty}^{\infty}\exp(-ikp)I_0dp + c_2\int_{-\infty}^{\infty}p\exp(-ikp)I_0 dp\right) \\
		+O(\epsilon^2),
	\end{gathered}
\end{equation}
reducing the problem at $O(\epsilon)$ to that of a Fourier transform, where we have
\begin{equation}
	\begin{gathered}
		c_1 = (z-h_0)\left( (\alpha^2\sin^2(\theta) + \beta^2\cos^2(\theta))\right) + \cos(\theta)a\beta^2 + ik\alpha^2\beta^2\tan(\theta) \\
		c_2 = \cos(\theta)\sin(\theta)(\alpha^2-\beta^2),
	\end{gathered}
\end{equation}
grouping the coefficients of each integral in variable $p$. Note that $c_1$ has $z$-dependence, which will become relevant later. For the moment, both $(x,z)$ are fixed. Now we complete the square inside of the integrand of all three integrals, bringing the extra factors of $\exp(-ikp)$ into the argument of the $\exp()$ in the second two, which will allow some simplification. We identify
\begin{equation}
	\begin{gathered}
		A = \frac{\cos^2(\theta)}{2\beta^2} + \frac{\sin^2(\theta)}{2\alpha^2} \\
		B = \frac{(z-h_0)\sin(\theta)\cos(\theta)}{\beta^2} - \frac{\sin(\theta)\left((z-h_0)\cos(\theta) + a\right)}{\alpha^2} + ik \\
		\tilde{B} = B-ik \\
		C = \frac{(z-h_0)^2\sin^2(\theta)}{2\beta^2} + \frac{\left((z-h_0)\cos(\theta) + a\right)^2}{2\alpha^2},
	\end{gathered}
\end{equation}
so that we may express 
\begin{equation}
	I_0 = \exp\left(\frac{\tilde{B}^2}{4A} - C\right)\exp\left(-\Big[p\sqrt{A} + \frac{\tilde{B}}{2\sqrt{A}}\Big]^2\right).
\end{equation}
We will then incorporate $-ikp$ into $I_0$ prior to evaluation. Consider that
\begin{equation}
\begin{gathered}
	\exp(-ikp)I_0 \\
	= \exp(-ikp)\exp\left(\frac{\tilde{B}^2}{4A}-C\right) \exp\left(-\Big[p\sqrt{A} + \frac{\tilde{B}}{2\sqrt{A}}\Big]^2\right) \\
	= \exp\left(\frac{\tilde{B}^2}{4A}-C\right)\exp\left(-\Big[p^2A + p(\tilde{B}+ik) + \frac{\tilde{B}^2}{4A}\Big]\right) \\
	= \exp\left(\frac{\tilde{B}^2}{4A}-C\right)\exp\left(-\Big[p^2A + pB + \frac{B^2}{4A}\Big]\right)\exp\left(\frac{k^2 + 2iBk}{4A}\right) \\
	= \exp\left(\frac{B^2}{4A}-C\right)\exp\left(-\Big[p^2A + pB + \frac{B^2}{4A}\Big]\right) \\
	= \exp\left(\frac{B^2}{4A}-C\right)\exp\left(-\Big[p\sqrt{A} + \frac{B}{2\sqrt{A}}\Big]^2\right),
\end{gathered}	
\end{equation}
a fact which will enable us to conserve some effort. It soon becomes clear that
\begin{equation}
	\begin{gathered}
		\int_{-\infty}^{\infty}I_0dp = \frac{\sqrt{\pi}}{{\sqrt{A}}}\exp\left(\frac{\tilde{B}^2}{4A} - C\right), \\
		\int_{-\infty}^{\infty}\exp(-ikp)I_0dp = \frac{\sqrt{\pi}}{{\sqrt{A}}}\exp\left(\frac{B^2}{4A} - C\right), \\
		\int_{-\infty}^{\infty}p\exp(-ikp)I_0dp = -\frac{B\sqrt{\pi}}{2A^{3/2}}\exp\left(\frac{B^2}{4A} - C\right)
	\end{gathered}
\end{equation}
This simplifies the resulting expressions and introduces a pleasant symmetry. Finally, we have a closed-form expression after carrying out the integration in $p$:
\begin{equation}
	\begin{gathered}
		P(x,z) = \frac{\cos(\theta)}{2\pi\alpha\beta}\left[\frac{\sqrt{\pi}}{{\sqrt{A}}}\exp\left(\frac{\tilde{B}^2}{4A} - C\right)\right] + \\
		\epsilon \tilde{h}_1\exp(ikx + \sigma t) \frac{\cos(\theta)}{2\sqrt{\pi A} \alpha^3 \beta^3} \exp\left(\frac{B^2}{4A} - C\right) \left(c_1 - \frac{Bc_2}{2A} \right) \\
		+O(\epsilon^2),
	\end{gathered}
\end{equation}
which will enable the matching of mean power deposition for an arbitrary Gaussian ellipsoid to stress modification within the amorphous bulk. We remind the reader that both the argument of the exponential and the term $\left( c_1 - \frac{Bc_2}{2A}\right)$ retain $z$-dependence, will be important because the above $P(x,z)$ will provide $\tau_0(z)$ and $\tau_{\epsilon}(z)$, or $\alpha_{10}(z)$ and $\alpha_{1\epsilon}(z)$. These expressions must then be integrated according to Equations (\ref{dispnreln1})-(\ref{dispnreln2}), which leads to the final linear stability result associated with the Gaussian ellipse characterized by the triplet $(a,\alpha,\beta)$.

\paragraph{Intuition: leading order.} 
Some simplification reveals that
\begin{equation}
	\exp\Bigg(\frac{\tilde{B}^2}{4A}-C\Bigg) = \exp\Bigg(-\frac{(z - h_0 + a\cos(\theta))^2}{2(\alpha^2\cos^2(\theta) + \beta^2\sin^2(\theta))  }   \Bigg).
\end{equation}
Thus the leading-order term is fundamentally a \textit{univariate} Gaussian about the $z$ axis whose angle-dependent standard deviation continuously interpolates between the nominal downbeam and crossbeam standard deviations at $0^{\circ}$ and $90^{\circ}$ respectively. The mean of this leading-order Gaussian is simply $h_0 - a\cos(\theta)$, which quite naturally says that the location of the region of highest power deposition rises to the surface as $\theta \to 90^{\circ}$. For mean $\mu$ and standard deviation $\sigma$ we have
\begin{equation}
\begin{gathered}
	\mu(\theta;a,\alpha,\beta) = h_0 - a\cos(\theta), \\
	\sigma(\theta;a,\alpha,\beta) = \sqrt{\alpha^2\cos^2(\theta) + \beta^2\sin^2(\theta)}
\end{gathered}
\end{equation}
by comparison with the usual form of a normal distribution.

\subsubsection{Integral evaluations}
We identify
\begin{equation}
	P_{0} = \frac{\cos(\theta)}{2\pi\alpha\beta}\left[\frac{\sqrt{\pi}}{{\sqrt{A}}}\exp\left(\frac{\tilde{B}^2}{4A} - C\right)\right],
\end{equation}
the leading-order part of the depth-dependence extracted from the bivariate Gaussian intensity field. We will also identify
\begin{equation}
	P_{1} = \frac{\cos(\theta)}{2\sqrt{\pi A} \alpha^3 \beta^3} \exp\left(\frac{B^2}{4A} - C\right) \left(c_1 - \frac{Bc_2}{2A} \right),
\end{equation}
the $O(\epsilon)$ part of the depth-dependence of the stress tensor following the Fourier transform of the surface in the small-perturbation limit. From the functional form of the dispersion relation described earlier, we will be required to carry out one integral of $P_0$ and the first through third integrals of $P_1$. While we could, in principle, write numerical evaluation of these integrals into the script that will later generate our theoretical predictions for $\theta_c$, this could possibly slow down our script considerably and introduce numerical errors.

\paragraph{$P_1$ evaluations.} It is useful to rewrite $P_1$:
\begin{equation}
\begin{gathered}
\frac{\cos(\theta)}{2\sqrt{\pi A}\alpha^3\beta^3}	\exp\left(R - \frac{Q^2}{4P} \right)\exp\left(\left[(z-h_0)\sqrt{P} + \frac{Q}{2\sqrt{P}}\right]^2\right)\left(E(z-h_0)+F\right),
\end{gathered}
\end{equation}
where constants (with respect to $z$) $E$ and $F$ are from the rearrangement of $c_1 - \frac{Bc_2}{A}$, so that we have
\begin{equation}
\begin{gathered}
	E = \alpha^2\sin^2(\theta) + \beta^2\cos^2(\theta) - \frac{c_2}{2A}\left( \frac{\sin(\theta)\cos(\theta)}{\beta^2} - \frac{\sin(\theta)\cos(\theta)}{\alpha^2}\right) \\
	F = \frac{c_2}{2A}\left( \sin(\theta)\frac{a}{\alpha^2} - ik\right) +  a\beta^2\cos(\theta) + ik\alpha^2\beta^2\tan(\theta)
\end{gathered}
\end{equation}
and
\begin{equation}
	\begin{gathered}
P=\frac{1}{4A}\left(\frac{\sin^2(\theta)\cos^2(\theta)}{\beta^4} - \frac{2\sin^2(\theta)\cos^2(\theta)}{\beta^2\alpha^2} + \frac{\sin^2(\theta)\cos^2(\theta)}{\alpha^4}\right) - \left(\frac{\sin^2(\theta)}{2\beta^2} + \frac{\cos^2(\theta)}{2\alpha^2}\right) \\
Q = \frac{1}{4A}\left(-2\sin^2(\theta)\cos(\theta)\frac{a}{\beta^2\alpha^2} + 2\sin(\theta)\cos(\theta)\frac{ik}{\beta^2} - \frac{2ik}{\alpha^2}\sin(\theta)\cos(\theta) + 2a\frac{\sin^2(\theta)\cos(\theta)}{\alpha^4}\right) - \frac{a\cos(\theta)}{\alpha^2} \\
R = \frac{1}{4A}\left( \frac{a^2\sin^2(\theta)}{\alpha^4} -2ik\sin(\theta)\frac{a}{\alpha^2} - k^2\right) - \frac{a^2}{2\alpha^2}
	\end{gathered}
\end{equation}
so that the problem is reduced to multiple integrations of the product of a(n affine) linear function and a Gaussian. We will define
\begin{equation}
	\begin{gathered}
		\tilde{P} = P\Bigg|_{k=0}, 	\tilde{Q} = Q\Bigg|_{k=0}, 	\tilde{R} = R\Bigg|_{k=0}, 	\tilde{F} = F\Bigg|_{k=0}
	\end{gathered}
\end{equation}
here for later use.

\paragraph{$P_0$ evaluation.}
Similarly, it is useful to rewrite $P_0$ prior to evaluation. We take
\begin{equation}
	\begin{gathered}
		P_0 = \frac{\cos(\theta)}{2\sqrt{\pi A}\alpha\beta}\exp\left(\frac{\tilde{B}^2}{4A} - C \right) = \\
		\frac{\cos(\theta)}{2\sqrt{\pi A}\alpha\beta}\exp\left(\tilde{R} - \frac{\tilde{Q}^2}{4\tilde{P}} \right)\exp\left(\left[(z-h_0)\sqrt{\tilde{P}} + \frac{\tilde{Q}}{2\sqrt{\tilde{P}}}\right]^2\right)
	\end{gathered}
\end{equation}
so that the problem is reduced to a single integration of a Gaussian.

\paragraph{Expression in physical parameters.} We re-express $P,Q,R,E,F$ in terms of basic physical quantities $\theta,a,\alpha,\beta$. We find
\begin{equation}
\begin{gathered}
	P = \frac{s^2c^2\alpha^2\beta^2}{2[\alpha^2c^2+\beta^2s^2]}\left[\frac{1}{\beta^2}-\frac{1}{\alpha^2}\right]^2 - \left[\frac{s^2}{2\beta^2} + \frac{c^2}{2\alpha^2}\right]; \\
	Q = \frac{-sc\alpha^2\beta^2}{\alpha^2c^2 + \beta^2s^2}\left[\frac{1}{\beta^2}-\frac{1}{\alpha^2}\right]\left[\frac{sa}{\alpha^2} - ik\right] - \frac{ac}{\alpha^2}; \\
	R = \frac{\alpha^2\beta^2}{2[\alpha^2c^2 + \beta^2s^2]}\left[\frac{sa}{\alpha^2} - ik\right]^2 - \frac{a^2}{2\alpha^2}; \\
	E = \left[\alpha^2s^2 + \beta^2c^2\right] - \left[ \frac{c^2s^2\alpha^2\beta^2(\alpha^2-\beta^2)}{\alpha^2c^2+\beta^2s^2}\right]\left[\frac{1}{\beta^2}-\frac{1}{\alpha^2}\right]; \\
	F = \frac{cs\alpha^2\beta^2(\alpha^2-\beta^2)}{\alpha^2c^2 + \beta^2s^2}\left[\frac{sa}{\alpha^2} - ik\right] + a\beta^2c + ik\alpha^2\beta^2\tan(\theta),
\end{gathered}
\end{equation}
where $s$ denotes $\sin(\theta)$ and $c$ denotes $\cos(\theta)$. As before, the ``$\sim$" version of a quantity denotes that $k=0$ there.

\subsubsection{Linear stability result}
\paragraph{Anisotropic plastic flow.} We find
\begin{equation}
	\text{Re}(\sigma_{APF}) = \frac{-3fA\cos(\theta)}{64A^{3/2}\sqrt{\pi}\tilde{P}^{7/2}\alpha^5\beta^5}\Big(\cos(2\theta)\omega_1 - \sin(2\theta)\omega_2\Big)k^2 + O(k^4),
\end{equation}
where
\begin{equation}
\begin{gathered}
	\omega_1 = \\ 2Ae^{\tilde{R}}\alpha^2\beta^2\Bigg[4h_0^2\tilde{P}^2\Big[2\sqrt{\tilde{P}}\left(E - e^{h_0(h_0\tilde{P} - \tilde{Q})}E + 2\tilde{P}\alpha^2\beta^2\right) + \sqrt{\pi}e^{-\frac{\tilde{Q}^2}{4\tilde{P}}}\Big(2\tilde{F}{\tilde{P}} - E\tilde{Q}\Big)\Big(\text{erfi}(\frac{2h_0\tilde{P}-\tilde{Q}}{2\sqrt{\tilde{P}}}) + \text{erfi}(\frac{\tilde{Q}}{2\sqrt{\tilde{P}}})\Big)\Big] \\ +  e^{\frac{-\tilde{Q}^2}{4\tilde{P}}}\Big[2e^{\frac{\tilde{Q}(-4h_0\tilde{P} + \tilde{Q})}{4\tilde{P}}}\sqrt{\tilde{P}}\Big(e^{\tilde{Q}h_0}(4E\tilde{P} + 2\tilde{F}\tilde{P} - E\tilde{Q}^2) + e^{\tilde{P}h_0^2}(-2\tilde{F}\tilde{P}(2\tilde{P}h_0 + \tilde{Q}) + E(4\tilde{P}(-1 + h_0^2\tilde{P} ) + 2h_0\tilde{P}\tilde{Q} + \tilde{Q}^2)  )    \Big) \\
	+ \sqrt{\pi}\Big(2\tilde{F}\tilde{P}(2\tilde{P}-\tilde{Q}^2) + E\tilde{Q}(\tilde{Q}^2-6\tilde{P})\Big)\Big(\text{erfi}(\frac{2\tilde{P}h_0 - \tilde{Q}}{2\sqrt{\tilde{P}}}) + \text{erfi}(\frac{ \tilde{Q}}{2\sqrt{\tilde{P}}})\Big) \Big]  \Bigg]
\end{gathered}
\end{equation}
and
\begin{equation}
\begin{gathered}
	\omega_2 =
	\Big[32Ae^{\tilde{P}h_0^2 - \tilde{Q}h_0 + \tilde{R}}x_0h_0\tilde{P}^{7/2}\alpha^4\beta^4\Big]
	-e^{\frac{-\tilde{Q}^2}{4\tilde{P}} + \tilde{R}}\Bigg[ \Big[2a\tilde{P}\beta^2 + \tilde{Q}(\alpha^2-\beta^2)c\Big]\times \\
	\Big[2e^{\frac{\tilde{Q}(-4\tilde{P}h_0 + \tilde{Q})}{4\tilde{P}}}\sqrt{\tilde{P}}\Big(e^{\tilde{Q}h_0}(-2\tilde{F}\tilde{P} + E\tilde{Q}) + e^{\tilde{P}h_0^2}(2\tilde{F}\tilde{P} - E(2\tilde{P}h_0 + \tilde{Q}))\Big) \\ + \sqrt{\pi}\Big[2E\tilde{P} + 2\tilde{F}\tilde{P}\tilde{Q} - E\tilde{Q}^2\Big]\Big[\text{erfi}\Big(\frac{2\tilde{P}h_0-\tilde{Q}}{2\sqrt{\tilde{P}}}\Big) + \text{erfi}\Big(\frac{\tilde{Q}}{2\sqrt{\tilde{P}}}\Big) \Big]\Big]\sin(\theta) \\
	+ 2\sqrt{\tilde{P}}\Bigg[\Big(e^{\frac{(-2h_0\tilde{P} + \tilde{Q})^2}{4\tilde{P}}}-e^{\frac{\tilde{Q}^2}{4\tilde{P}}}\Big)
	\Big((2E\tilde{P} + 2\tilde{F}\tilde{P}\tilde{Q} - E\tilde{Q}^2)(\alpha^2-\beta^2)cs\Big) \\
	+ 2\sqrt{\pi}\sqrt{\tilde{P}} \Big(\text{erfi}\Big(\frac{2\tilde{P}h_0-\tilde{Q}}{2\sqrt{\tilde{P}}}\Big)
	+ \text{erfi}\Big(\frac{\tilde{Q}}{2\sqrt{\tilde{P}}}\Big)\Big)\Big(-(\tilde{F}\tilde{P} - E\tilde{Q})(\alpha^2-\beta^2)cs + \tilde{P}\tilde{Q}\alpha^2\beta^2(c_2 -2A\alpha^2\beta^2\tan(\theta))\Big) \\
	+ e^{\frac{\tilde{Q}(-4h_0\tilde{P} + \tilde{Q})}{4\tilde{P}}} \Bigg(\Big(2h_0\tilde{P} - \tilde{Q}\Big)\Big(e^{\tilde{Q}h_0}(E\tilde{Q}-2\tilde{F}\tilde{P}) + e^{\tilde{P}h_0^2}\Big(2\tilde{F}\tilde{P} - E(2\tilde{P}h_0 + \tilde{Q}) \Big)\Big)(\alpha^2-\beta^2)cs \\
	+ 2\tilde{P}\Big[\Big(e^{\tilde{P}h_0^2}E - e^{\tilde{Q}h_0}(E -2\tilde{F}\tilde{P}h_0 + E\tilde{Q}h_0)\Big)\Big(\alpha^2-\beta^2\Big)cs + 2\Big(e^{\tilde{P}h_0^2} - e^{\tilde{Q}h_0}\Big)\tilde{P}\alpha^2\beta^2\Big(c_2 - 2A\alpha^2\beta^2\tan(\theta)\Big)\Big]\Bigg)\Bigg] \Bigg].
\end{gathered}
\end{equation}

\paragraph{Ion-induced swelling.} We find
\begin{equation}
	\text{Re}(\sigma_{IIS}) = -\frac{\hat{\alpha}\cos(\theta)}{64(\sqrt{A}\sqrt{\pi}\tilde{P}^{7/2}\alpha^3\beta^3)}\exp\left(-\tilde{Q}h_0 - \frac{\tilde{Q}^2}{4\tilde{P}} + \tilde{R}\right)\omega_3k^2 + O(k^4)
\end{equation}
where
\begin{equation}
	\begin{gathered}
		\omega_3 = -2\exp\Big(\frac{\tilde{Q}^2}{4\tilde{P}} \Big)\sqrt{\tilde{P}}\Big[e^{\tilde{P}h_0^2}\Big(2\tilde{F}\tilde{P}(2h_0\tilde{P} + \tilde{Q}) + E(4h_0^2\tilde{P}^2 - \tilde{Q}^2 + \tilde{P}(4-2h_0\tilde{Q}))\Big) + 8h_0^2\tilde{P}^3\alpha^2\beta^2  \\
		+ e^{\tilde{Q}h_0}\Big[E(-4\tilde{P} - 8h_0^2\tilde{P}^2 + \tilde{Q}^2) - 2(\tilde{F}\tilde{P}\tilde{Q} + 8h_0^2\tilde{P}^3\alpha^2\beta^2)\Big]\Big] \\
		+ \sqrt{\pi}e^{\tilde{Q}h_0}\left(2\tilde{F}\tilde{P}(2\tilde{P} + 8h_0^2\tilde{P}^2 - \tilde{Q}^2) + E\tilde{Q}(-6\tilde{P} - 8h_0^2\tilde{P}^2 + \tilde{Q}^2)\right)\Big(\text{erfi}(\frac{2h_0\tilde{P}-\tilde{Q}}{2\sqrt{\tilde{P}}}) + \text{erfi}(\frac{\tilde{Q}}{2\sqrt{\tilde{P}}}      )\Big)       
	\end{gathered}
\end{equation}

\paragraph{Composite.} In the limit of small cross-terms, which we have determined to be physical and appropriate, the linear dispersion relation is simply 
\begin{equation} \label{composite}
\text{Re}(\sigma) = \text{Re}(\sigma_{IIS}) + \text{Re}(\sigma_{APF}).
\end{equation}
We will find that this approximation is sufficient for our purposes at this time: the determination of $\theta_c$, the minimal $\theta > 0$ such that Equation (\ref{composite}) becomes positive, signaling the onset of pattern formation. We note that in the long-wave limit, IIS is negative for all $\theta$, contributing stabilization of the surface (or, equivalently, contributing to the decay of the perturbation). This is well-aligned with its description as a stabilization mechanism, albeit in the uniform strength case, in \cite{Swenson_2018,evans-norris-JPCM-2022}.

\section{Results}
\subsection{Interface relation and spatial variation} 

\paragraph{Film thickness estimation.}

\begin{figure}[h!]
	\centering
	\includegraphics[totalheight=5cm]{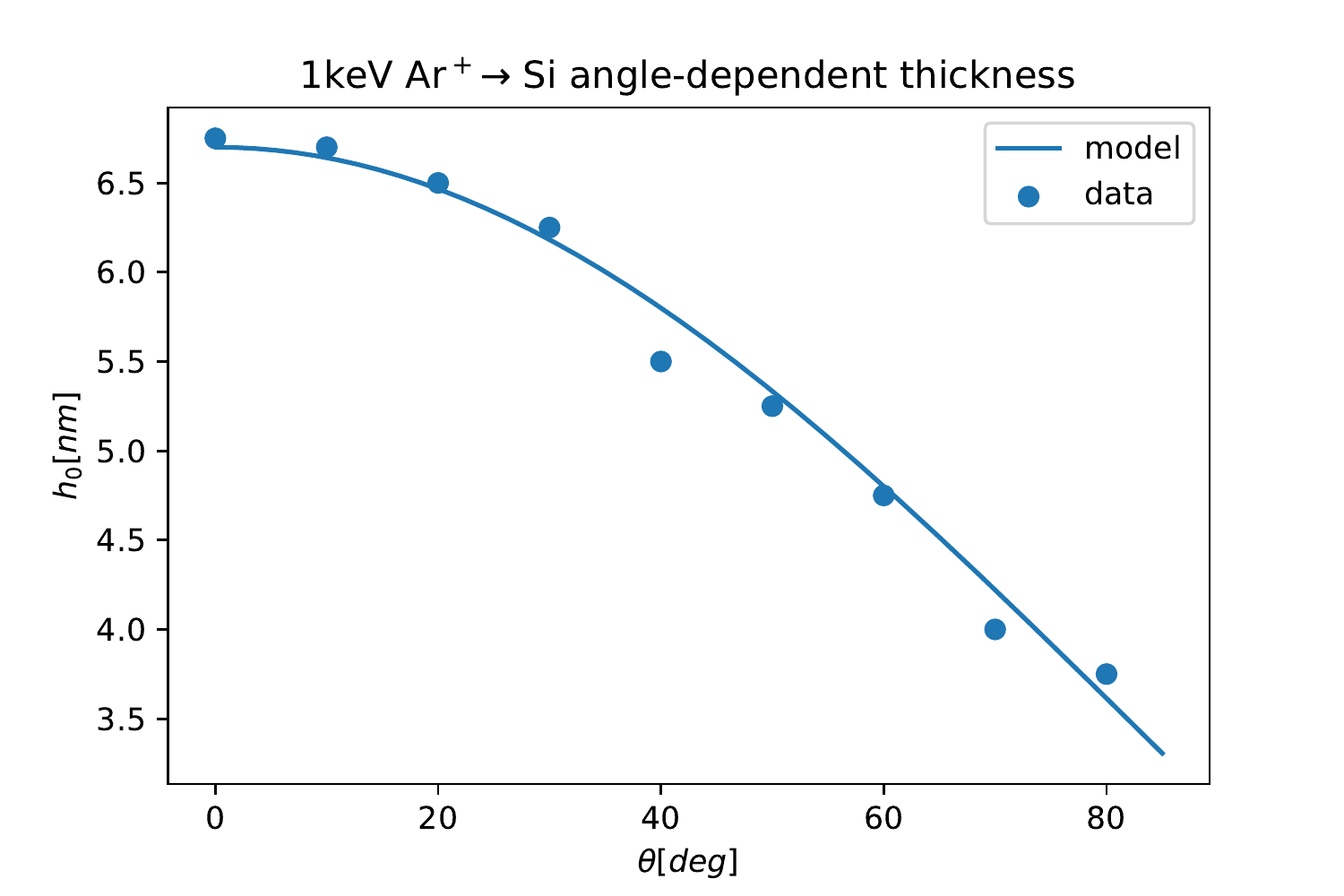} 
	\includegraphics[totalheight=5cm]{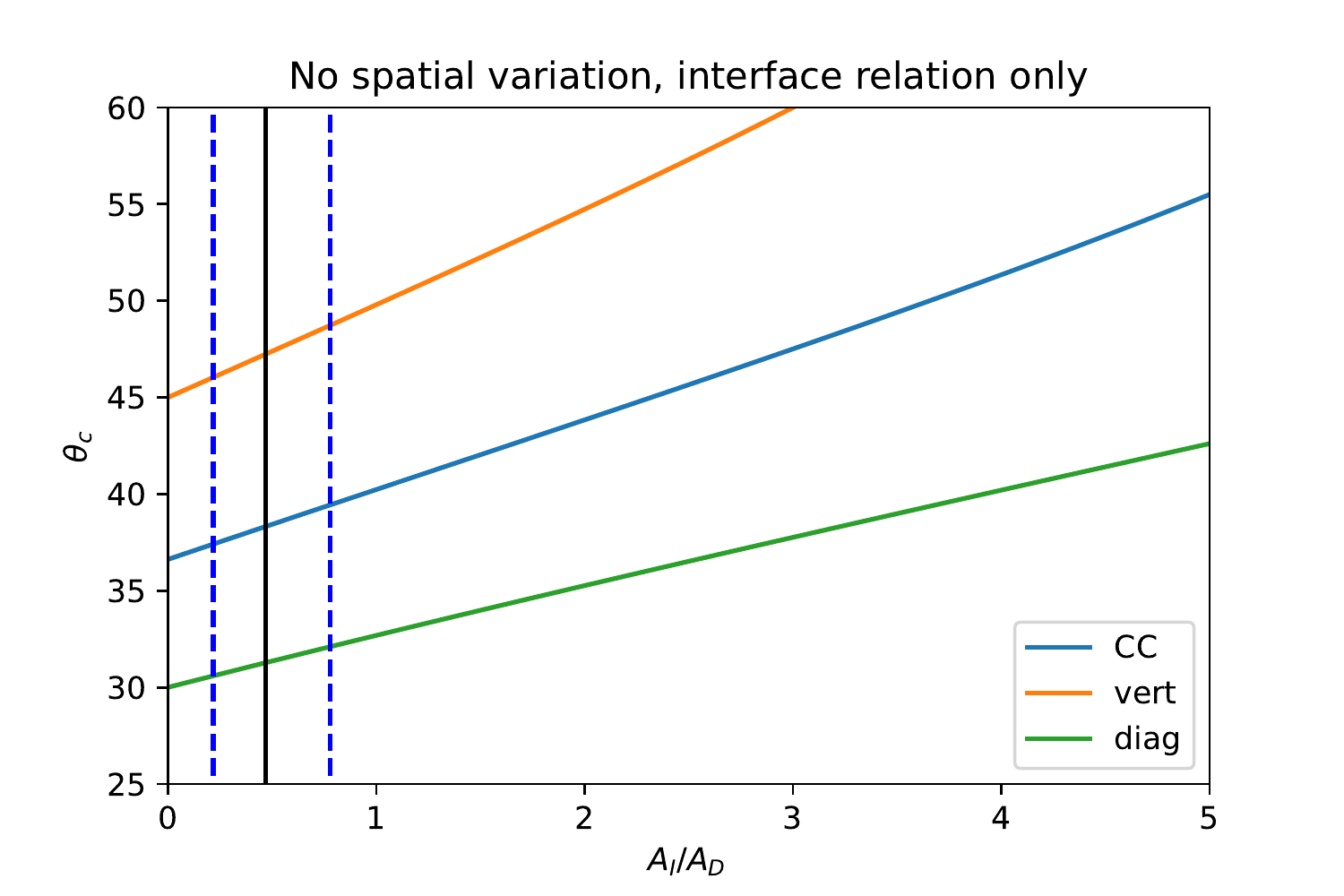}
	\caption{\textbf{Left:} Angle-dependent film thickness predicted by the present model compared with data generated by fitting GISAXS data and TRI3DST \cite{norris-etal-SREP-2017}. \textbf{Right:} $\theta_c$ against $\frac{A_I}{A_D}$ for three different interface relations with no spatial variation in ion-induced stress components throughout the bulk and no flux dilution, in order to emphasize the influence of the interface relations themselves. As expected in our previous work, the physically-grounded interface relation produces $\theta_c$ somewhere in between the two opposite extremes. Vertical lines are values of $\frac{A_I}{A_D}$ and the error bounds computed later in the text. \textbf{``CC"} denotes that the interfaces were generated according to the present model based on the collision cascade. \textbf{``Vert"} denotes that the interfaces were assumed to be vertical translations. \textbf{``Diag"} denotes that the interfaces were assumed to phase shift as $\propto \sin(\theta)$ and thin as $\propto \cos(\theta)$. }
	\label{fig:thicknessandthetac}
\end{figure}

As discussed previously elsewhere \cite{moreno-barrado-etal-PRB-2015}, it is not immediately obvious what the correct notion of the lower interface should be. One idea described elsewhere in the prompt-regime context is to use the mean downbeam deposition depth plus two standard deviations of ion implantation from the center of a Gaussian ellipsoid \cite{sigmund-PR-1969,sigmund-JMS-1973,hoffsass-etal-APA-2012,norris-etal-SREP-2017}. Based on this line of reasoning, we have compared our theoretical film depth from Equation \ref{thicknesseqn} against the angle-dependent film depth as computed by matching to GISAXS data \cite{norris-etal-SREP-2017}. It is clear from Figure \ref{fig:thicknessandthetac} that there is very strong agreement between these quasi-empirical results and our newly-derived closed-form expression. This implies the suitability of the present interface relation for use in our analysis and elsewhere. While there is not much angle-dependent film-thickness data available, we do note that our $h_0(\theta)$ expression appears to agree with Figure 2 of \cite{moreno-barrado-etal-PRB-2015} for 500eV Xe$^+$ on Si.

\paragraph{Direct influence on $\theta_c$-selection. } 
In our previous work, we observed that the assumption of vertically-applied stress and vertically-translated interfaces, or diagonally-applied stress and diagonally-translated interfaces, caused any depth-dependence imposed along those axes to simply ``drop out" of the dispersion relation, leading to no effect on $\theta_c$. Observing that the vertical case would produce a higher value of $\theta_c$ than the diagonal case given the same ratio $\frac{A_I}{A_D}$, we were motivated to refine the interface relation. Here, we briefly consider the comparison of the vertical case, diagonal case, and the case of the collision-cascade interface relation with the assumption of constant beam-induced stress components throughout the bulk.

\paragraph{Steady-state stress vs MD and BCA.} Our steady-state stress distribution due to the Gaussian-ellipsoidal power deposition compares well against the MD simulations of \cite{moreno-barrado-etal-PRB-2015} for 300eV Xe$^+ \to$ Si at $60^{\circ}$ and the distribution generated by our own Monte Carlo simulation. We obtain an estimate of peak power deposition at about 1.8nm below the surface and that half-peak power occurs at about 1nm below the surface, in agreement with their work. It is also noteworthy that \cite{steinbach-etal-PRB-2011} has associated nuclear stopping power with isotropic swelling previously, albeit for high energies.

\begin{figure}[h!]
	\centering
	\includegraphics[totalheight=6cm]{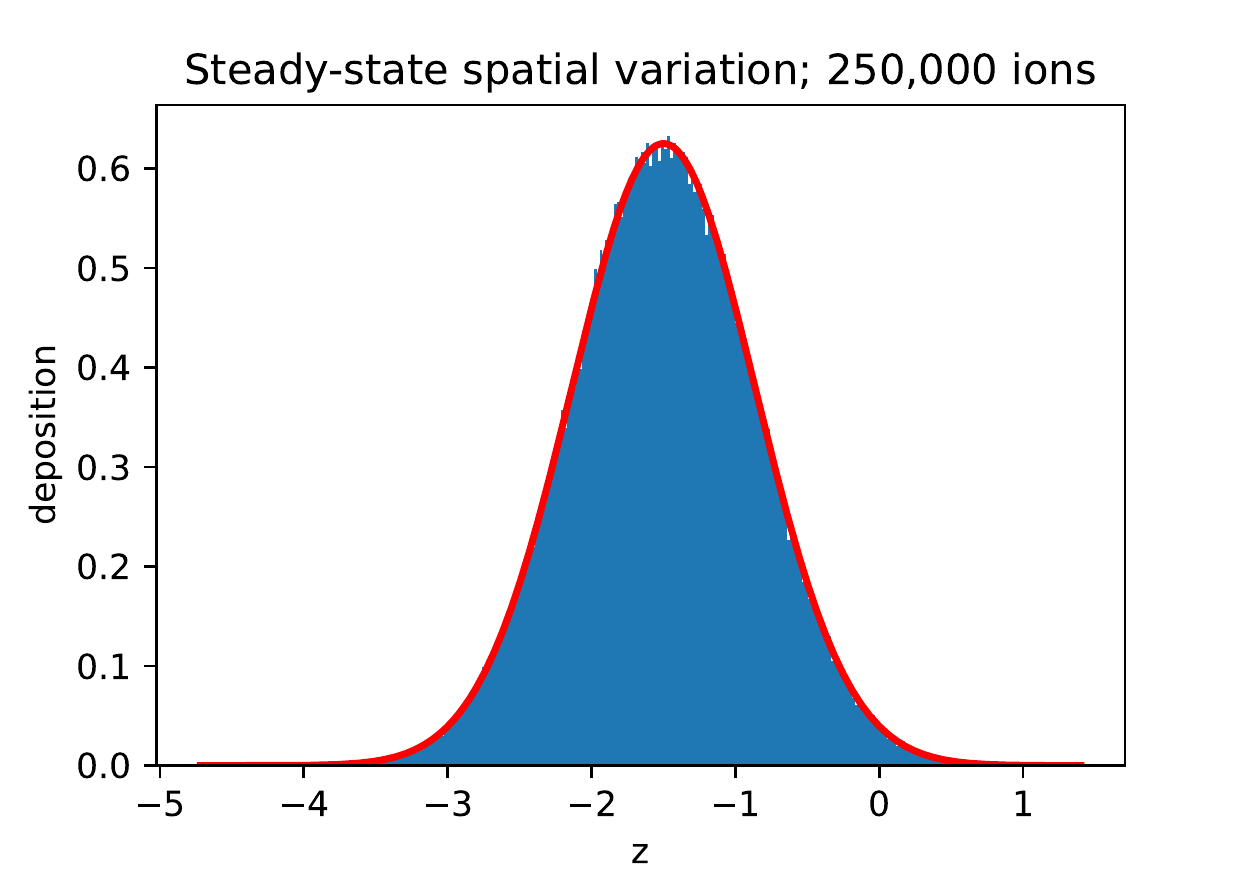}
	\caption{Monte Carlo vs theorized steady-state stress for 300eV Xe$^+ \to$ Si based on parameters obtained from BCA. Here, height ``0" is taken as the flat upper interface, with depth below this level taken as negative.}
\end{figure}

\paragraph{Notation for parameters.}
Throughout, we have used $fA$ as a coefficient of the APF-term and $\hat{\alpha}$ as a coefficient of the IIS term. Both have natural units of $\frac{1}{\text{s}}$, being, fundamentally, \textit{rates}. As has been suggested elsewhere \cite{ishii-etal-JMR-2014,norris-etal-SREP-2017,NorrisAziz_predictivemodel}, it is natural to consider these quantities as fundamentally flux-dependent. Hence we will consider
\begin{equation}
	fA \to fA_{D}, \hspace{.25cm} \hat{\alpha} \to fA_{I},
\end{equation}
with subscripts denoting the deviatoric and isotropic parts of deformation respectively. This makes the assumption that both terms scale with flux. While this seems probable, too little is known about the underlying physics at the present time. We therefore adopt this notation on a tentative basis.

\paragraph{Implications for parameter estimation: 250 eV Ar$^+ \to$ Si.}
In the present work, we have modified our previous assumptions about the form of the depth-dependence of stress and our modeling of film thickness. It is therefore appropriate to incorporate these adjustments into our parameter estimates. We note that our steady-state depth-dependence is of the form
\begin{equation}
\tau(z;0,h_0) = \alpha_1(z;0,h_0) = \frac{\cos(\theta)}{2\sqrt{\pi A}\alpha\beta}\exp\left(\frac{-(z-h_0+a\cos(\theta))^2}{2(\alpha^2\cos^2(\theta) + \beta^2\sin^2(\theta))}   \right),
\end{equation}
where the $\cos(\theta)$ originates from flux dilution along an unperturbed, flat free interface. Now computing the mean about the film depth, we obtain
\begin{equation} \label{steadystatestress}
\begin{gathered}
	<\tau(z;0,h_0)> \hspace{.25cm} = \hspace{.25cm} <\alpha_1(z;0,h_0)> \hspace{.25cm} = \\ \frac{\cos(\theta)}{2h_0}\left[ \text{erf}\left(\frac{a\cos(\theta)}{\sqrt{2}\sqrt{\alpha^2\cos^2(\theta)+\beta^2\sin^2(\theta)}}\right) - \text{erf}\left(\frac{a\cos(\theta)-h_0}{\sqrt{2}\sqrt{\alpha^2\cos^2(\theta)+\beta^2\sin^2(\theta)}}\right) \right],
\end{gathered}
\end{equation}
a substantially more complicated form than that of our previous work. As before, we will cancel the $\cos(\theta)$ prefactor with the $\cos(\theta)$ of the biaxial modulus for the purposes of comparison with stress data. Here, we take
\begin{equation}
\begin{gathered}
h_0 = a\cos(\theta) + 2\left(\sqrt{\alpha^2\cos^2(\theta) + \beta^2\sin^2(\theta)}\right),
\end{gathered}
\end{equation}
and re-fit parameters $A_I$ and $A_D$ to the experimental data of Perkinson \cite{perkinsonthesis2017}. We fit the mean steady-state in-plane stress,
\begin{equation}
	\begin{gathered}
		<T_{0}^{11}> = -2fA\eta(D_{11}-D_{33})<\tau(z;0,h_0)> - 2\hat{\alpha}\eta<\alpha_1(z;0,h_0)>  \\ = \\ -6fA\eta\cos(2\theta)<\tau(z;0,h_0)> - 2\hat{\alpha}\eta<\alpha_1(z;0,h_0)>,
	\end{gathered}
\end{equation}
with these assumptions, leading to the estimates
\begin{equation}
	\begin{gathered}
		fA\eta \approx 0.2449 \pm 0.0255 \text{ GPa} \\
		\hat{\alpha}\eta \approx 0.1148 \pm 0.0561. \text{ GPa}
	\end{gathered}
\end{equation}
There are two significant implications of these estimates. First, using the flux $f = 1.2\times 10^{1} \frac{\text{ions}}{\text{nm}^2\cdot \text{s}}$ (which was the same as that of \cite{madi-thesis-2011}) and the value $\eta \approx 1.5\times 10^2 \text{ Gpa} \cdot \text{s}$ from \cite{norris-etal-SREP-2017}, this leads to the estimates
\begin{equation}
	fA \approx 1.632 \times 10^{-3} \frac{1}{\text{s}},
\end{equation}
as compared with the estimate $fA \approx 3 \times 10^{-4} \frac{1}{\text{s}}$ for 1keV Ar$^{+} \to$ Si \cite{norris-etal-SREP-2017}, and
\begin{equation}
	\begin{gathered}
		A_D \approx 1.36\times 10^{-4} \frac{\text{nm}^2}{\text{ion}} \\ 
		A_I \approx 6.37\times 10^{-5} \frac{\text{nm}^2}{\text{ion}}.
	\end{gathered}
\end{equation}

\begin{figure}[h!]
	\centering
	\includegraphics[totalheight=6cm]{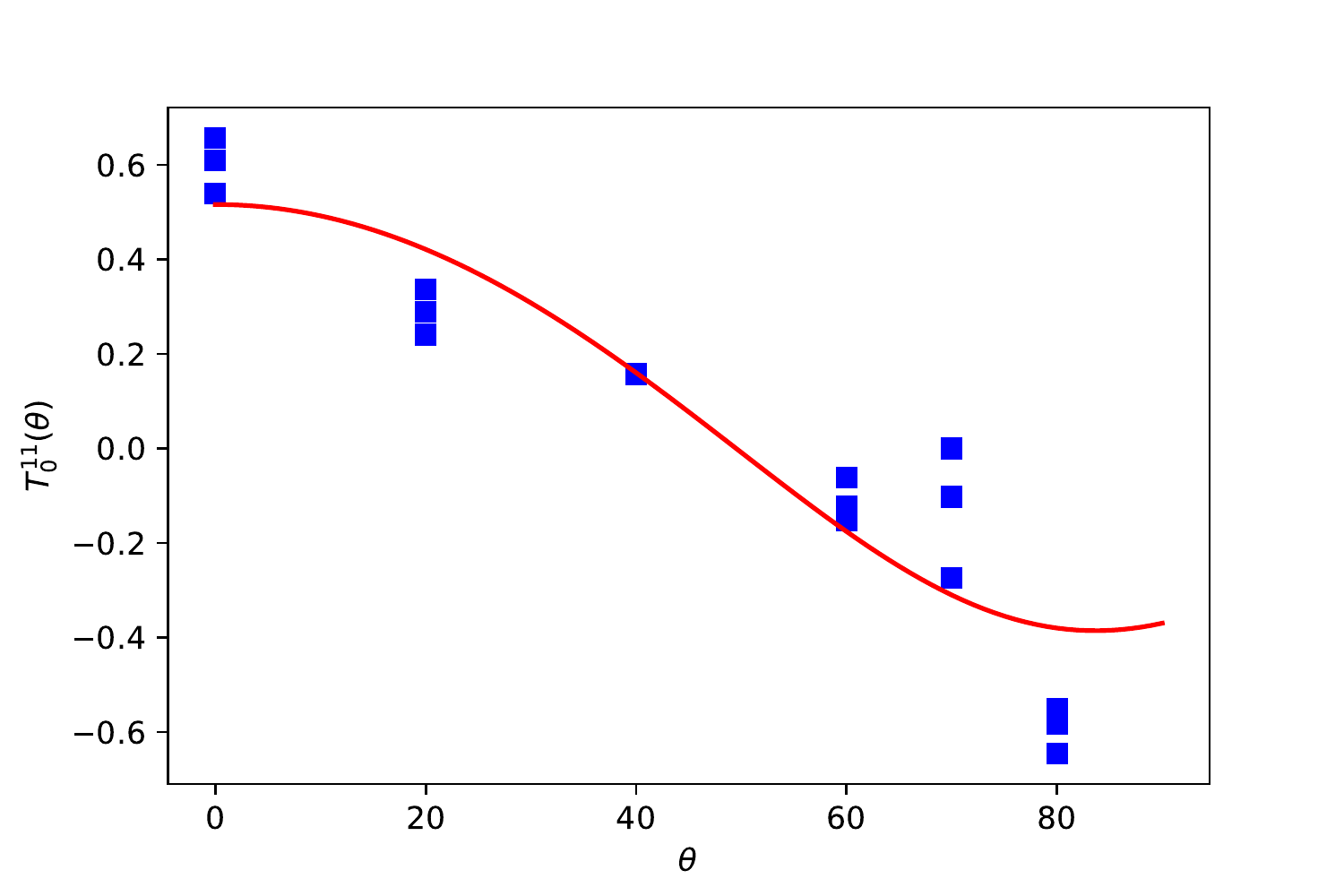}
	\caption{Comparison of fitted values and experimental stress measurements, backstepped with a corrected film thickness.}
\end{figure}
We note that these values are within about an order of magnitude of the reported values from \cite{norris-etal-SREP-2017} and \cite{madi-thesis-2011}. The second important implication is that these values lead to a substantially smaller estimate of the ratio $\frac{A_I}{A_D}$ than we had previously obtained, which will tend to suggest a smaller theoretical prediction for $\theta_c$---unless the refinement of the interface relation turns out to exert a strongly stabilizing effect.

\paragraph{Implications for wavelength predictions from full-spectrum model.} 
We take a moment to consider the implications of the interface relation for a full-spectrum composite model including APF and IIS \textit{in the absence of depth-dependence}, as our analysis has been conducted in the long-wave limit only; a full-spectrum analysis will be the subject of future work and is expected to be of greater complexity than that of the present work. Nonetheless, such an effort will be of significant value if the long-wave theory is capable of correctly identifying $\theta_c$ (thereby satisfying one of two minimal conditions for a unifying theory of ion-induced pattern formation.)

Previously, with the assumption of vertically-translated interfaces and a parameter estimate for $fA\eta$, apparent good agreement had already been obtained using a model without any swelling \cite{norris-PRB-2012-linear-viscous}. However, we have shown that the assumptions made about interfacial geometry can be extremely influential in $\theta_c$-selection. It is natural to hypothesize that such an extreme influence would likewise be seen in wavelength selection. Although we do not currently have a full-spectrum analysis for arbitrary depth-dependence profiles as we have for long-wave perturbations in the present work, there is enough data and modeling to make a very preliminary investigation.

We wish to to obtain wavelength predictions from our experimental estimates for parameter groups $fA_D\eta$ and $fA_I\eta$ with angle-dependent film thickness and the shape of the collision cascade taken into consideration. Towards this end, and with permission from the author, we rework a part of the derivation of the full-spectrum linear dispersion relation of \cite{norris-PRB-2012-linear-viscous} and generalize its treatment of the lower interface to include the interface relation term $\frac{\tilde{g}_1}{\tilde{h}_1}$. This calculation is shown in the Appendix and leads to only a very slight modification of the originally-reported result. Using the resulting full-spectrum linear dispersion relation \textit{with mechanism strength held uniform across the film}, we then numerically determine the most unstable wavenumber at each irradiation angle, which leads to the theoretical prediction of which wavelength should be experimentally observed at each angle. 
\begin{figure}[h!]
	\centering
	\includegraphics[totalheight=6.5cm]{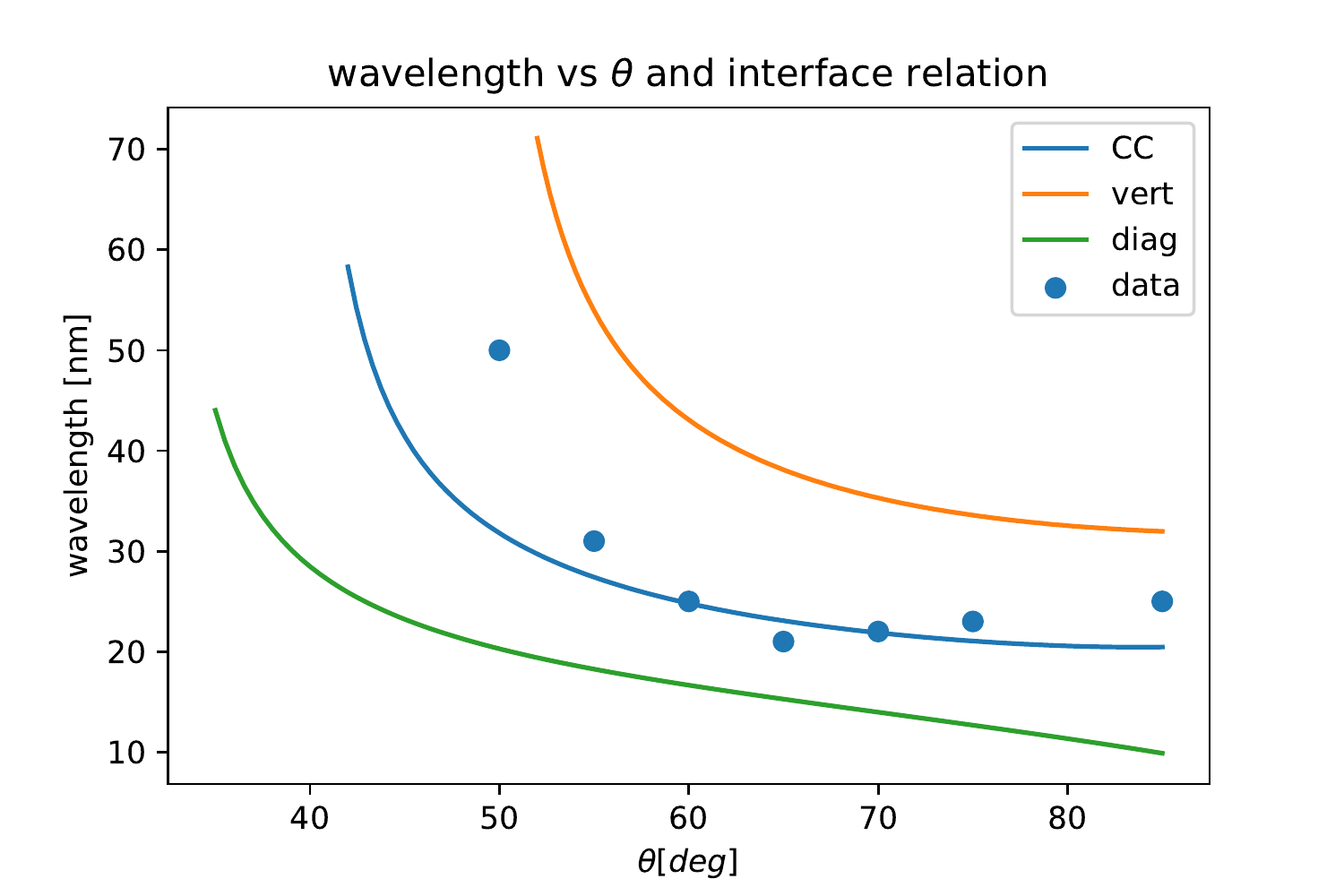}
	\caption{A look at the influence of the interface relation and parameter estimates on wavelength selection. Using parameter estimates for $fA_D\eta$ and $fA_I\eta$ obtained from downbeam Legendre polynomial depth dependence or for uniform stress. \textbf{``CC"} denotes that the interfaces were generated according to the present model based on the collision cascade. \textbf{``Vert"} denotes that the interfaces were assumed to be vertical translations. \textbf{``Diag"} denotes that the interfaces were assumed to phase shift as $\propto \sin(\theta)$ and thin as $\propto \cos(\theta)$. \textbf{``Data"} is from experimental measurements and was obtained from \cite{madi-etal-2008-PRL,madi-etal-PRL-2011,madi-aziz-ASS-2012}. }
\end{figure}
We find that the combination of the interface relation determined from BCA parameters in simulations of 250eV Ar$^+ \to$ Si and experimental estimates according to our previous work lead to strong agreement between the experimental, angle-dependent wavelength data of \cite{madi-thesis-2011} and our theoretical predictions. This preliminary check of the full-spectrum version of our analysis constitutes the \textit{third} experimental data to be at least qualitatively described by our present model, alongside angle-dependent film thickness and in-plane stress.

In calculating these wavelengths with the assumption of uniform strength of APF and IIS across the film, we have used the earlier estimates from our previous work. There, parameter estimates were obtained by assuming spatial variation according to Legendre polynomials about the downbeam direction. For steady-state in-plane stress averaged about the film depth, spatial variation of the ``downbeam-Legendre" type has no effect on the parameter estimates due to the integral-preserving property of Legendre polynomials. This is not true for spatial variation induced by computing power deposition according to Gaussian ellipsoids.  First, the steady-state in-plane stress \textit{will} depend on the details of these ellipsoids, which makes parameters obtained with such assumptions more sensitive to the details of spatial variation than those obtained by ``downbeam-Legendre", as the spatial variation is implicitly a part of the fitting. This is apparent when comparing the form of the in-plane stress from our previous work with that of Equation (\ref{steadystatestress}) of the present work. Intuitively, one may observe that under the ``downbeam Legendre" assumption, all power is deposited \textit{somewhere} in the film and flux dilution cancels, which eliminates one source of dependence on the geometry of the upper interface. However, in the case of spatially-resolved Gaussian ellipsoids, flux dilution is non-trivial, and power may be ``deposited", nonphysically, \textit{above} the amorphous bulk; these two factors lead to geometry-dependent stress modification not captured by spatial averaging. Allowing part of power-deposition to occur nonphysically above the bulk may lead to fitted parameter values that are higher than those that would be needed to experience the same extent of stabilization (or destabilization) were \textit{all} power deposited in the bulk. ``Re-averaging" and assigning those parameter estimates as representative of \textit{uniform} ion-induced stress modification across the film is then clearly inappropriate.

The indication of power deposition ``above the bulk" is a quirk of the Gaussian ellipsoid model and has been noted elsewhere \cite{hossain-etal-JAP-2012,hobler-etal-PRB-2016,NorrisAziz_predictivemodel}. Indeed, such a ``re-averaging" would obscure the well-known Bradley-Harper instability \cite{bradley-harper-JVST-1988,bradley-PRB-2011b}, as the instability is driven \textit{precisely} by spatial variation. However, such a model is in widespread use and is simple enough to allow analytical work, while nuanced enough to adequately captured relevant aspects of the physics. In many systems and for many incidence angles, it is a good approximation. Wavelength extraction in a full-spectrum version of the present spatially-resolved model is therefore a topic for future work, as is moving beyond a simple Gaussian ellipsoidal model for power deposition.

\subsection{Single-ellipsoid model: co-existing mechanisms}
\paragraph{Hypothesis formation.} As has been discussed elsewhere \cite{norris-PRB-2012-linear-viscous,van-dillen-etal-APL-2001-colloidal-ellipsoids,van-dillen-etal-APL-2003-colloidal-ellipsoids,van-dillen-etal-PRB-2005-viscoelastic-model}, there is no existing justification for the phenomenological use of the stress tensor \ref{APFtensor} in the nuclear stopping regime other than that it apparently produces good agreement with experimental and simulation results. The true physical explanation for the use of such a tensor, originally motivated by the electronic stopping regime's melt-cycle, is not currently known. A fundamental assumption in the present work is of the spatial variation of anisotropic plastic flow and isotropic swelling occur within the irradiated film: do they occur together, in the same regions of the film, or do they occur in a spatially-separated way, with swelling occurring in one part and anisotropic plastic flow occurring in another part; and with what distribution? Perhaps the most natural starting point is to assume, \textit{a priori}, that there is no spatial separation, and both APF and IIS originate within the collision cascade simultaneously, without maintaining any serious commitment to this idea.

\begin{figure}[h!]
	\centering
	\includegraphics[totalheight=5cm]{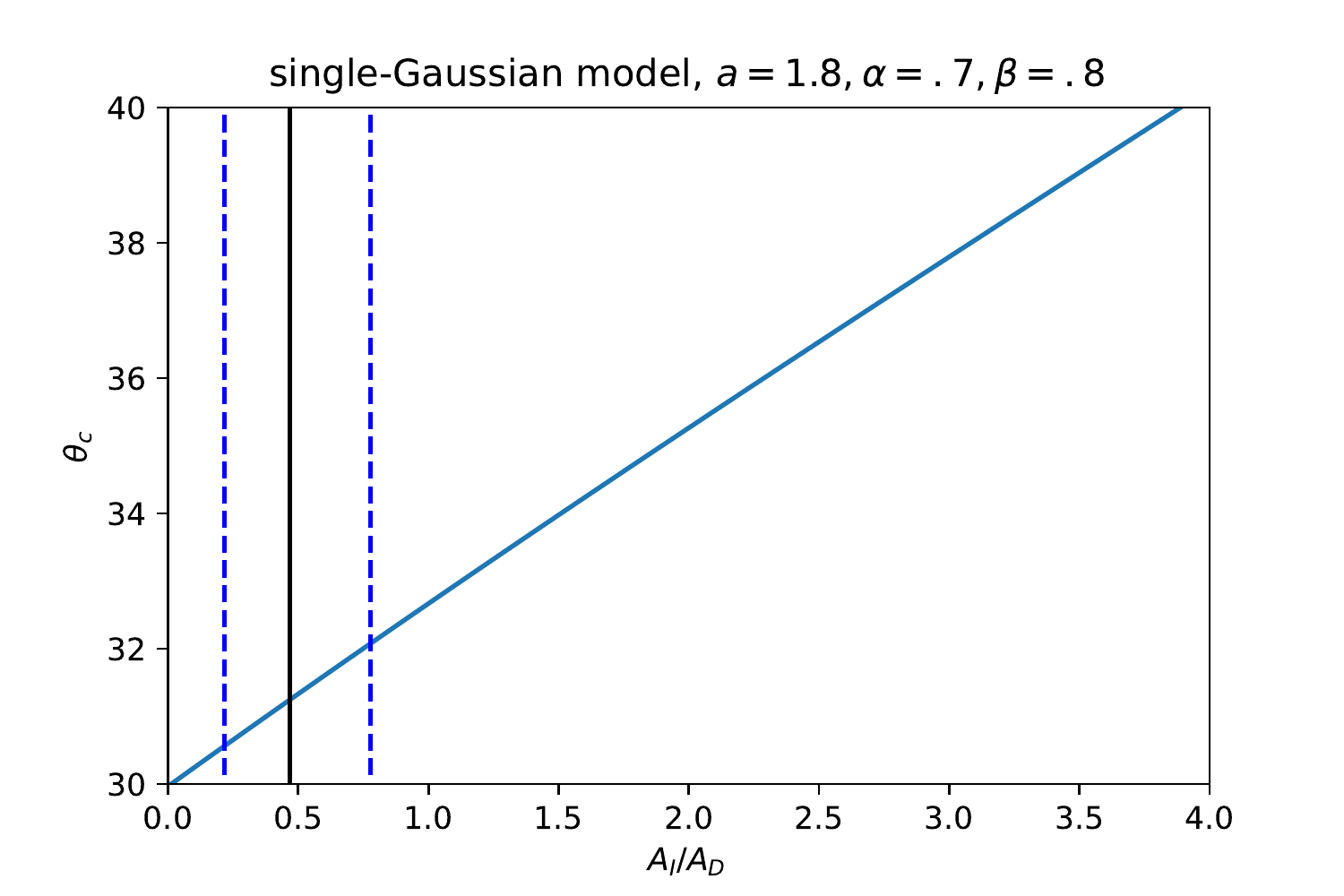}
	\includegraphics[totalheight=5cm]{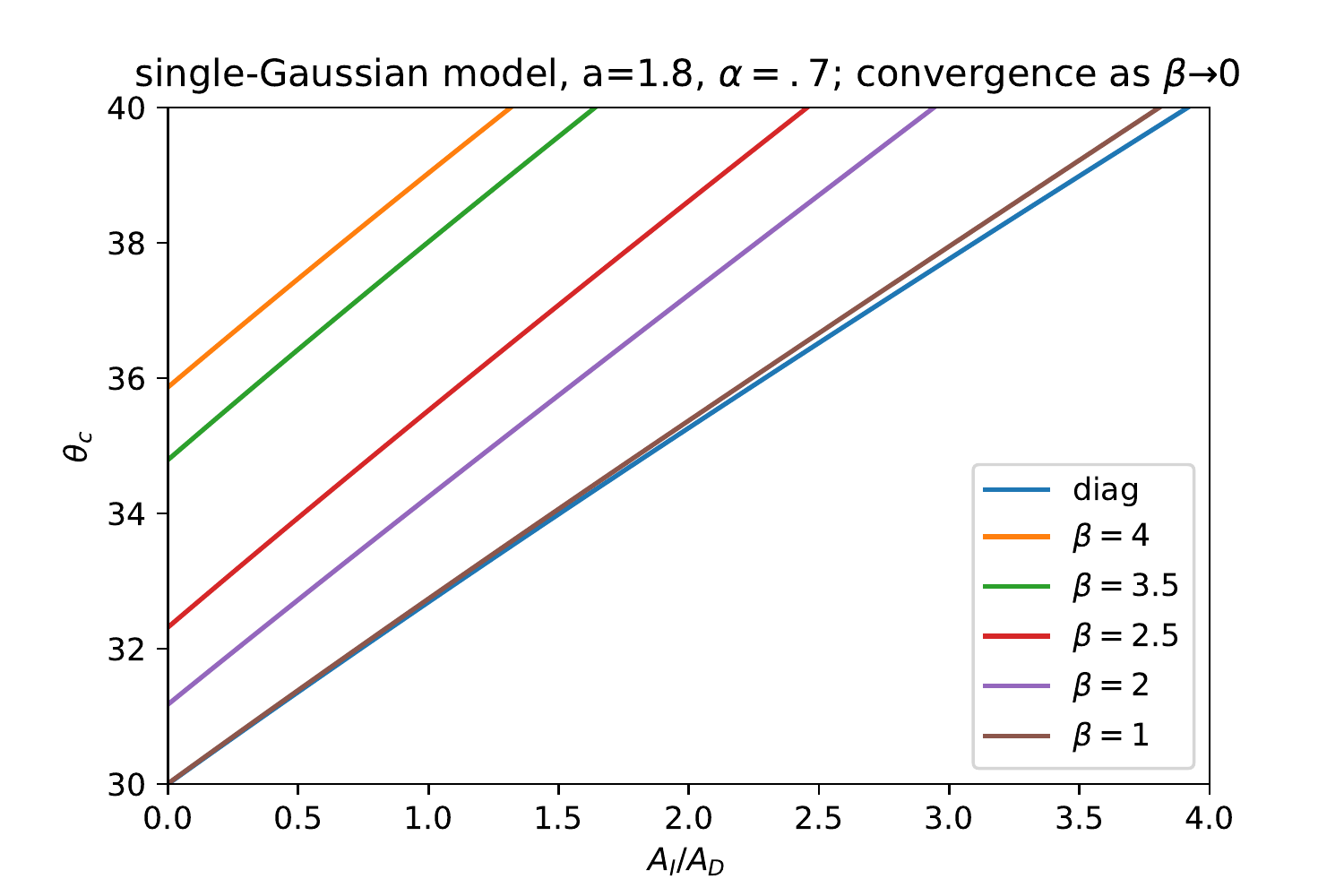}
	\caption{$\theta_c$ vs $\frac{A_I}{A_D}$ under the ``single-Gaussian" assumption with physical parameter values. \textbf{Left:} Critical angle predictions for the 250eV Ar$^+ \to$ Si system using BCA parameters and the single-ellipsoid model of spatial variation. \textbf{Right:} Demonstration of the convergence of the single-ellipsoid model to the predictions of the ``downbeam-Legendre", diagonal-translation model as cross-beam width vanishes. Interestingly, for small ratios $\frac{A_I}{A_D}$, this convergence is evidently rapid even for $\beta =1$; however, it is not uniform in $\frac{A_I}{A_D}$ and for large $\frac{A_I}{A_D}$, $\beta$ must be quite small to become indistinguishable from the pure diagonal ``downbeam-Legendre" case.}
	\label{fig:singlegaussian}
\end{figure}

\paragraph{Convergence of single-ellipsoid model to ``ion-tracks" as $\beta \to 0$.} In the way of consistency-checking the present, single-ellipsoid model, we verify that in the limit as crossbeam straggle $\beta \to 0$, the interface relation becomes that of our previous work and elsewhere \cite{moreno-barrado-etal-PRB-2015,Swenson_2018} and all power deposition is concentrated along the ion-track. This is shown in Figure \ref{fig:singlegaussian}, a plot of $\theta_c$ predictions for a=1.8, $\alpha=.7$ (the downbeam penetration and straggle parameters for 250eV Ar$^+ \to$ Si) with $\beta$ varying. As expected, for small crossbeam straggle, the ``diagonal ion-tracks" model approximates $\theta_c$ selection.

\paragraph{Critical angle selection: 250eV Ar$^+ \to$ Si.}
In combination with the parameter values estimated above, we find that using the parameter values for the resulting distribution of ions, $a=1.8, \alpha=.7, \beta=.8$ (associated with 250eV Ar$^+ \to$ Si) leads to a high-end estimate $\theta_c \approx 32^{\circ}$ and a low-end estimate $\theta_c \approx 30.5^{\circ}$. This result leads us to believe that there is something wrong with one of the assumptions that we have made up until this point. Because these predictions strongly disagree with experimentally-observed values of $\theta_c$ even though we have refined our treatment of the interface relation, which is, itself, in good agreement with other experimental results, our suspicion turns to the assumption of the form of the depth-dependence.

\subsection{Double-ellipsoid model: spatially-separated mechanisms}

\begin{figure}[h!]
	\centering
	\includegraphics[totalheight=4.5cm]{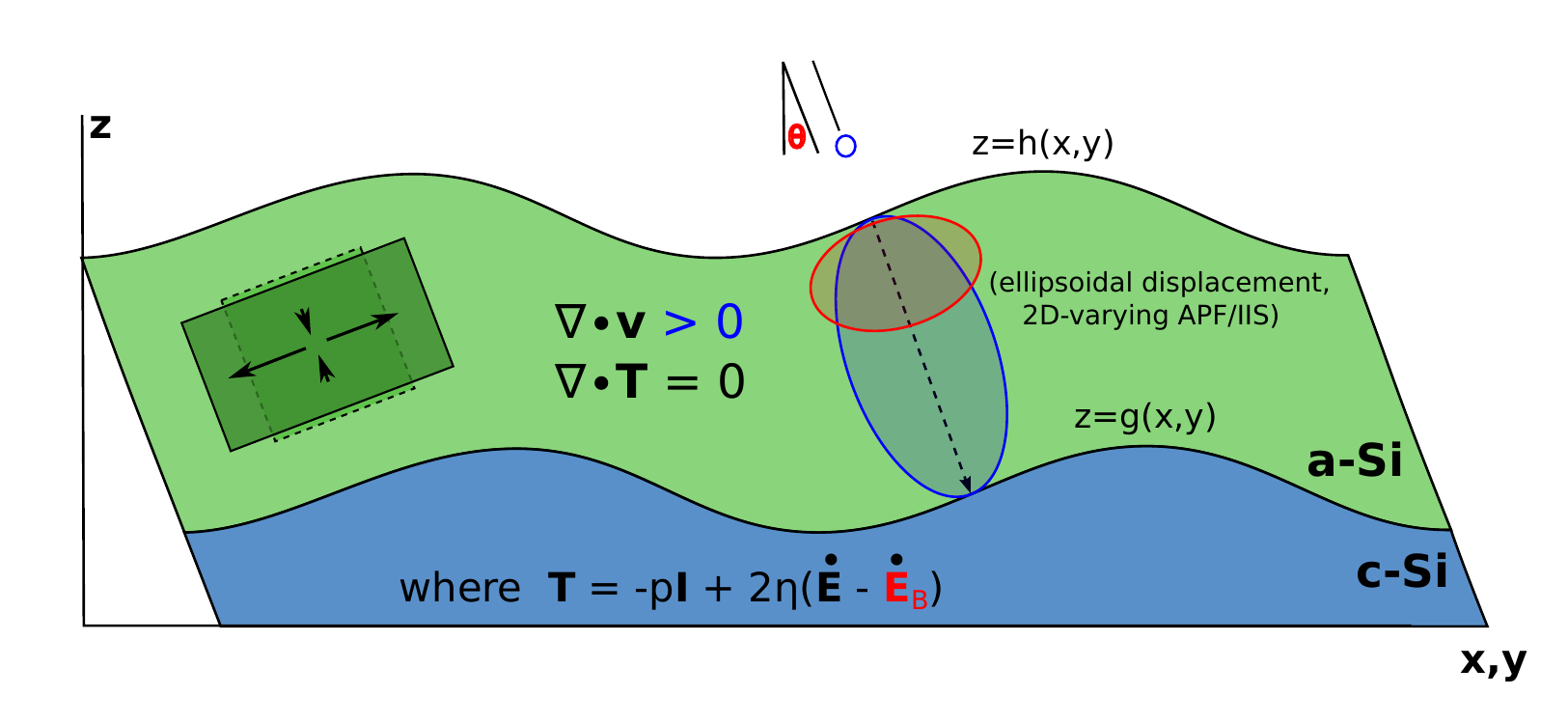}
	\caption{Schematic depicting ion bombardment at an incidence angle of $\theta$ and stress induced in the thin film by ion implantation. We hypothesize that IIS may develop primarily within the ion-implanted region, while APF may develop somewhere else within the collision cascade. This schematic depicts the hypothesis that APF originates from displacements ``up-beam" within the collision cascade, which we have modeled, for simplicity, as existing within its own Gaussian ellipsoid.}
	\label{fig:schematic3}
\end{figure}

\paragraph{Hypothesis formation.} Despite significant modifications to enhance the physical realism of our model, and a resulting new estimate of the values of $A_I$ and $A_D$, the ``co-existing mechanisms" model characterized by a single Gaussian ellipsoid fails \textit{spectacularly} to reproduce the experimentally-observed critical angle of $\theta_c \approx 48^{\circ}$ seen for 250eV Ar$^+ \to$ Si experiments. This is at first unsettling, because we have eliminated multiple degrees of freedom in our analysis, refining the interface relation and connecting the depth-dependence profiles of each mechanism directly to power deposition, all of which \textit{individually} seem to agree with known experimental data. A natural assumption might be that the models of isotropic swelling and anisotropic plastic flow are fundamentally incorrect, but they, together, seem to capture the observation of in-film volumization due to ion-implantation \cite{moreno-barrado-etal-PRB-2015} as well as the angle dependence of the mean in-plane stresses \cite{perkinsonthesis2017}. Furthermore, a combination of APF, viscous relaxation and erosion has lead to a highly parsimonious explanation of experimental data in \cite{norris-etal-SREP-2017}. Swelling, in particular, also seems to fit some of the trends seen in other ion-target-energy systems \cite{Teichmann2013,evans-norris-JPCM-2022}. Nuclear stopping has been associated with volumization elsewhere \cite{steinbach-etal-PRB-2011}, and at least some MD simulations appear to show stress profiles commensurate with isotropic swelling due to ion-implantation \cite{moreno-barrado-etal-PRB-2015}.

This suggests that the disagreement originates from the \textit{arrangement} of the model's parts rather than the parts individually. Because most previous approaches have essentially treated these mechanisms as averaged across the film depth, it is unsurprising that this possibility has gone unnoticed. The assumption of the physical placement of each mechanism within the film bulk is fundamental and we have, up until this point, made a na\"{i}ve first assumption. We now turn attention to the possibility of spatially-separated mechanisms, while noting that the hypothesis of spatially-separated mechanisms is actually the \textit{prevailing} hypothesis throughout the literature, even if it is not explicitly identified as such. The Bradley-Harper theory \cite{bradley-harper-JVST-1988,bradley-PRB-2011b}, like the crater function framework and related modifications (\cite{kalyanasundaram-etal-APL-2008,kalyanasundaram-etal-JPCM-2009,norris-etal-2009-JPCM,norris-etal-NIMB-2013,norris-arXiv-2014-pycraters} and many others), is fundamentally surface-oriented, originally purely erosive and redistributive in character with surface diffusion as a regularization mechanism. Later, the Orchard model of surface-confined viscous flow replaced surface diffusion in order to improve agreement with experimental results \cite{orchard-ASR-1962,umbach-etal-PRL-2001,norris-etal-SREP-2017}. These changes correspond to a trend of increasing ``hydrodynamicization" of modeling approaches as the bulk film behavior and stress evolution are increasingly believed to be of great importance for achieving a comprehensive, first-principles and predictive theory \cite{ishii-etal-JMR-2014,chan-chason-JVSTA-2008,perkinsonthesis2017,NorrisAziz_predictivemodel}.

Consequently, in the analysis and modeling of such systems, it has become unclear where the influence of surface mechanisms begins and the influence of bulk mechanisms ends, as these are fundamentally modeling choices rather than indications of the underlying physics. As was seen in \cite{hossain-etal-JAP-2012,hobler-etal-PRB-2016}, the erosion rate due to power deposition is actually nonlinear, despite frequently-used simplifying assumptions, and varies according to local density; this immediately implies complex interactions between erosion, bulk stress, and in-film defect dynamics of the type studied by \cite{chan-chason-JVSTA-2008,ishii-etal-JMR-2014}. In the work of \cite{norris-PRB-2012-linear-viscous}, where the stress tensor associated with anisotropic plastic flow was shown to lead to good agreement between experimental wavelength data and theoretical predictions, it has been discussed that anisotropic plastic flow, as seen from the film bulk, could possibly be understood as anisotropic stress generation due to surface redistribution. That is: the apparent phenomenon of anisotropic plastic flow could be due, in fact, to prompt, near-surface rearrangements caused by surface redistribution--- an effect neglected in purely stress-based hydrodynamic models such as ours. This is based in part upon the lack of an obvious mechanism in the nuclear stopping regime that would mirror the ``APF" stress tensor, originally based on the ``melt-cycle" phenomenon observed in the electronic stopping regime, while both APF and surface redistribution lead to coefficients $\propto \cos(2\theta)$ in the projected downbeam direction (x) and $\propto \cos^2(\theta)$ in the y-direction \cite{carter-vishnyakov-PRB-1996,norris-PRB-2012-linear-viscous,NorrisAziz_predictivemodel}. The current lack of clarity on this matter strongly motivates a unification of all bulk and surface effects into a single framework capable of addressing the interactions of stress, defects, ion-implantation, erosion, redistribution and viscous flow directly from a data-informed theory of power deposition in order to avoid over-counting, under-counting or misattributing mechanisms (e.g., gradual-regime APF versus prompt-regime redistribution).

We will now consider a model with two Gaussian ellipsoids: one for APF and one for IIS. We will take IIS to originate from the ion-implanted region of the nuclear collision cascade (i.e., ``final resting places" of the ions, as computed from SRIM \cite{ziegler-biersack-littmark-1985-SRIM}), motivated (1) by the agreement of BCA statistics and the construction of the steady-state depth-dependence with the stress generation reported for 300eV Xe$^+$ on Si in \cite{moreno-barrado-etal-PRB-2015}, which shows stresses associated with volumization exactly in the ion-implanted region; (2) discussion elsewhere \cite{trinkaus-ryazanov-PRL-1995-viscoelastic} that single-atom displacements outside of a narrow, cylindrical ion track are expected to be approximately spherical and therefore contribute only isotropic stresses; (3) that implantation of noble gas ions is already understood to lead to local changes in density at higher energies \cite{chini-etal-PRB-2003-TEM}; and (4) findings that nuclear stopping may lead to isotropic stresses and volumization in at least some cases \cite{steinbach-etal-PRB-2011}. In-film defect dynamics are also known to directly influence in-plane stress due to \cite{chan-chason-JVSTA-2008,ishii-etal-JMR-2014}, and the ion-implanated ellipsoid within the film is expected to be rich in such defects. We will then suppose that APF may possibly develop in some other region still characterized by a bivariate Gaussian in downbeam-crossbeam coordinates. While we do \textit{not} assert that APF insertion \textit{must} follow a bivariate Gaussian in reality, we will use this as a simple first model of a separate, APF-dominated region, while being relatively confident that most of the generation of ion-induced swelling should occur within the ion-implanted region.

\paragraph{Adaptive refitting of experimental parameters.}
Changing our hypothesis of where the individual mechanisms operate (i.e., their spatial variation) within the film will change our estimate of the experimental values $A_I$ and $A_D$.
For simplicity, we will assume that APF follows a deposition profile according to its own Gaussian ellipsoid, analogous to that of IIIS, but having parameter values $a_2, \alpha_2, \beta_2$ to distinguish them from the same Gaussian description assigned to IIS. By toggling these parameters, re-fitting parameter estimates from the experimental data for each, and generating new $\theta_c$ predictions, we may attempt a simple ``hypothesis testing". This will give some insight into how the spatial variation of APF, on its own, alters $\theta_c$-selection for 250eV Ar$^+ \to$ Si while we keep IIS fixed to BCA parameters $a=1.8, \alpha=.7, \beta=.8$ obtained via SRIM.

\paragraph{Critical angle selection: 250eV Ar$^+ \to$ Si with parameter refitting.}
\begin{figure}[h!]
	\centering
	\includegraphics[totalheight=4.5cm]{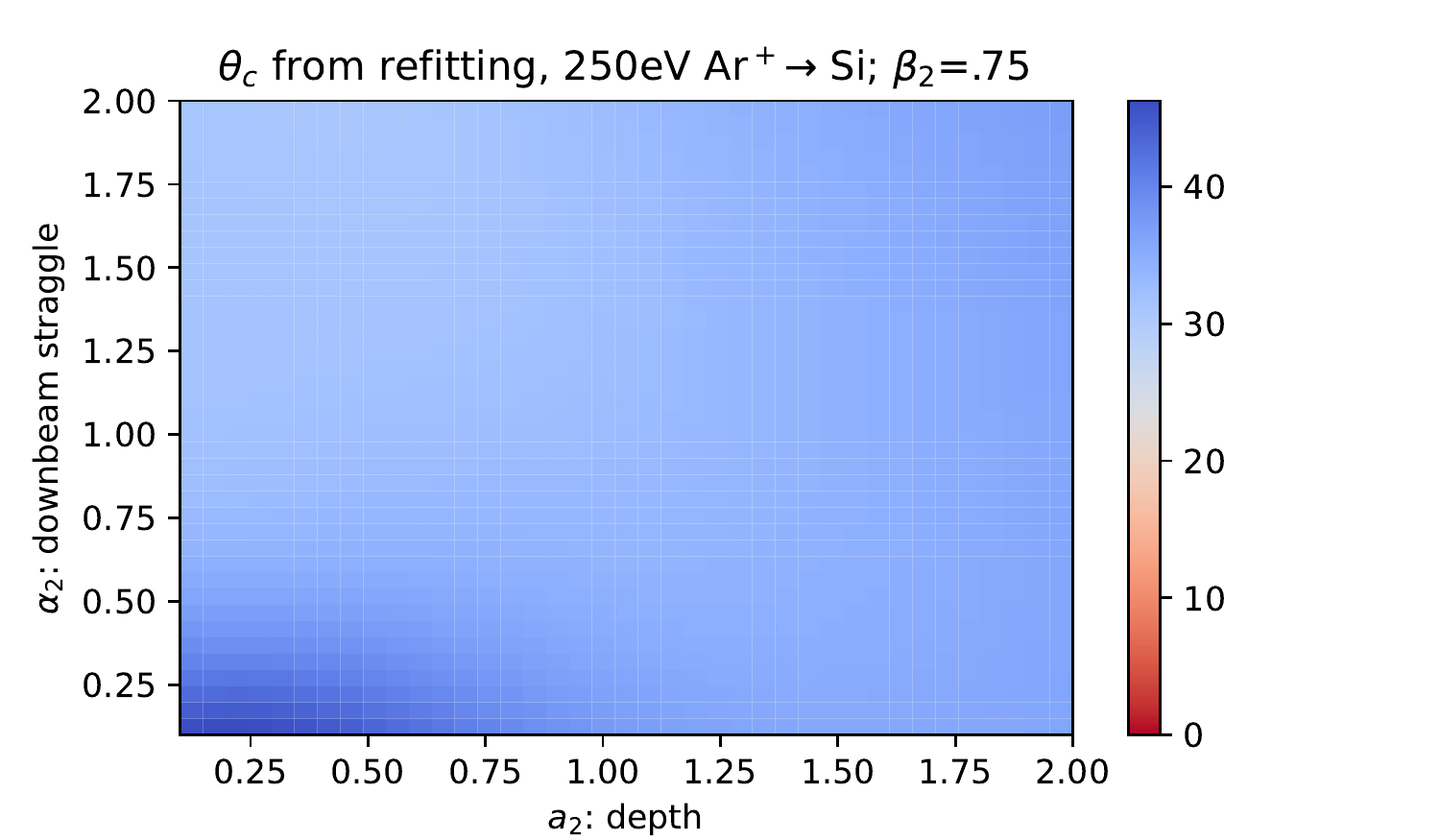}
	\includegraphics[totalheight=4.5cm]{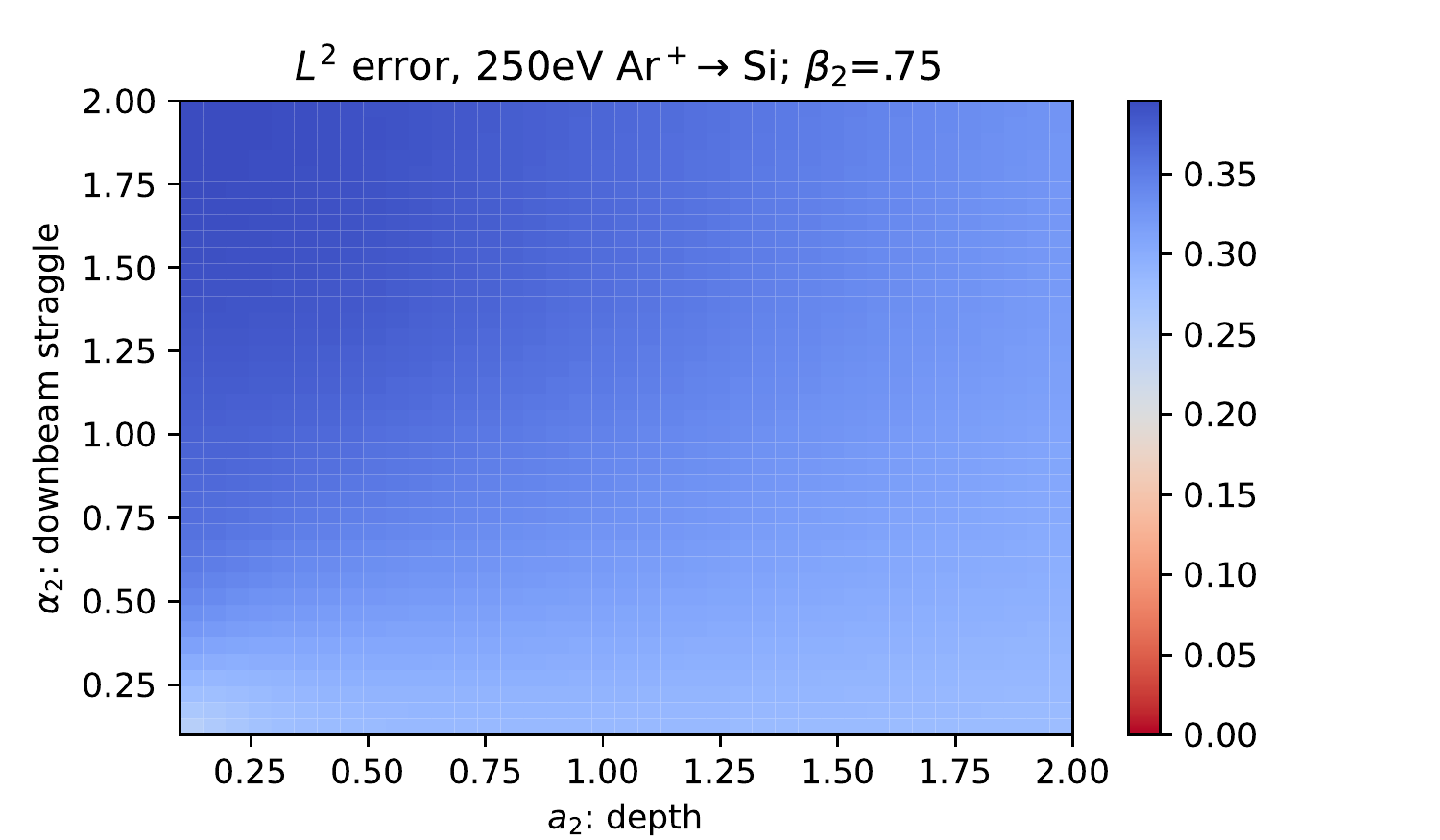}
	\includegraphics[totalheight=4.5cm]{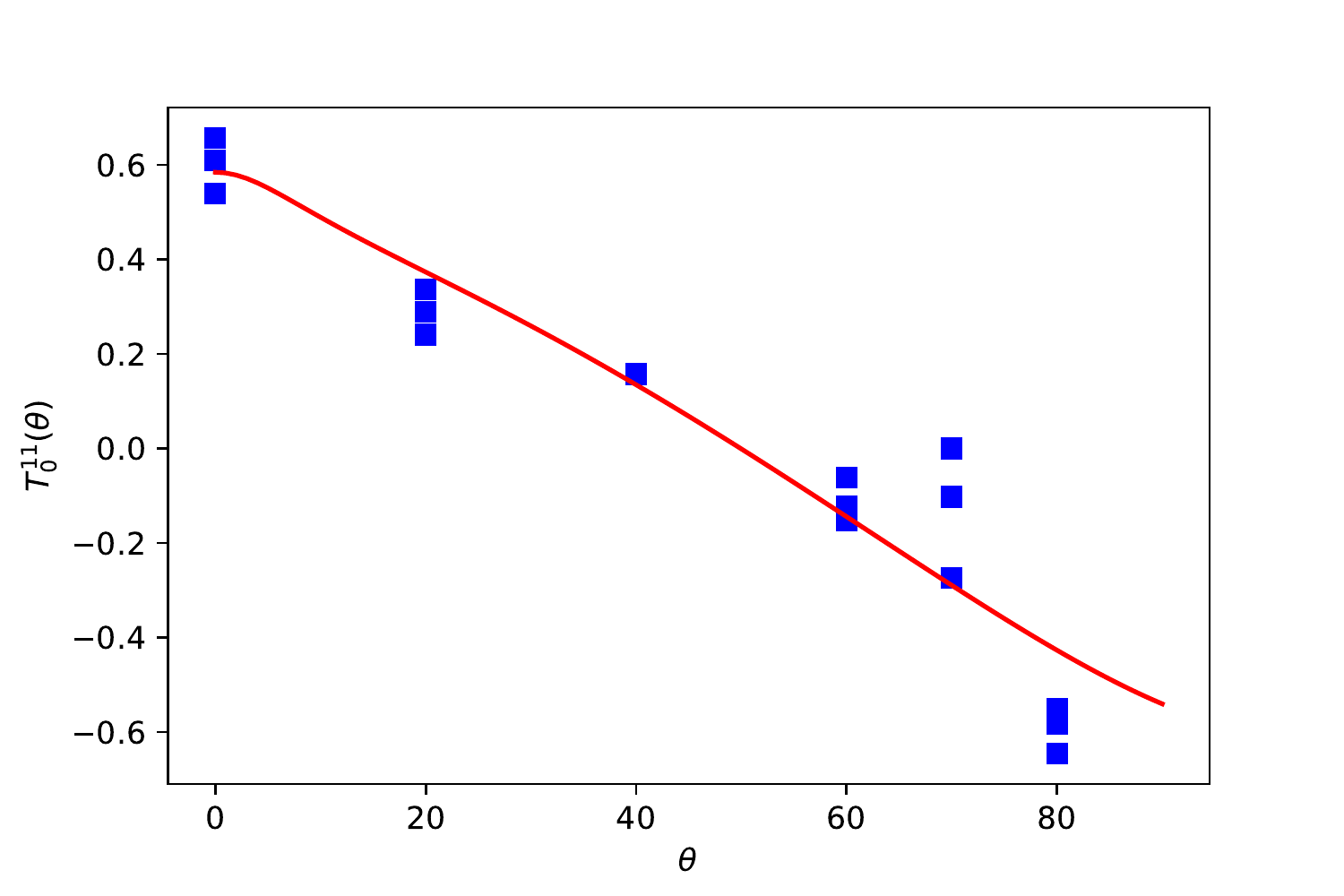}
	\caption{\textbf{Left:} $\theta_c$ predictions from adaptive re-fitting of $fA_I\eta$ and $fA_D\eta$ against Perkinson's stress data based on \textit{hypothesized} origination parameters for APF $a_2,\alpha_2,\beta_2$. IIS assumed to originate in the ion-implanted region of the collision cascade with $a=1.8, \alpha=.7, \beta=.8$. \textbf{Right:} $L^2$ error between the model's steady-state in-plane stress and that of the experimental data \cite{perkinsonthesis2017}. \textbf{Comment:} Interestingly, the parameter fits that minimize $L^2$ error also suggest that APF develops within an ellipsoid that is shallow in the downbeam coordinate and wide in the crossbeam coordinate. \textbf{Lower:} Comparison of fitted values and experimental stress measurements, backstepped with a corrected film thickness and generated from $L^2$ error minimization of the double-ellipsoid model.}
	\label{fig:heatmaps2}
\end{figure}

With the parameters $a,\alpha,\beta$ fixed for IIS and determined via SRIM, we allow $a_2,\alpha_2,\beta_2$ to vary for APF. We then minimize the $L^2$ error between the experimental data and our theoretical $T_0^{11}$, the in-plane stress component. Within our search domain $(a_2,\alpha_2,\beta_2) \in [.1,4]^3$, error is uniquely minimized when $a_2=.1,\alpha_2=.1,\beta_2 \approx .75$. For these parameters of APF's Gaussian ellipsoid, our model predicts $\theta_c \approx 46^{\circ}$ in the middle of the uncertainty range, in excellent agreement with the range $\theta_c \approx 45^{\circ}-48^{\circ}$ observed for 250eV-1keV Ar$^+ \to$ Si irradiation \cite{madi-etal-2008-PRL,madi-etal-PRL-2011,madi-aziz-ASS-2012,moreno-barrado-etal-PRB-2015,hofsass-bobes-zhang-JAP-2016,perkinsonthesis2017}. We also find parameter estimates
\begin{equation}
\begin{gathered}
	fA\eta \approx 0.3314 \pm 0.0270 \text{ GPa} \\
	\hat{\alpha}\eta \approx 0.1013 \pm 0.0450 \text{ GPa},
\end{gathered}
\end{equation}
which are comparable to those previously obtained. In Figure \ref{fig:heatmaps2}, we show heatmaps of $\theta_c$ and $L^2$ against the parameters for APF's Gaussian ellipsoid where $\beta_2 \approx .75$, the $L^2$ error-minimizing value. We also show a comparison of the theoretical and experimental angle-dependent in-plane stresses. The smallest $\theta_c$ given by the uncertainties in the parameters $fA\eta$ and $\hat{\alpha}$ is $\sim44.3^{\circ}$, while the largest is $\sim47.9^{\circ}$. This is, again, in excellent agreement with the range of experimental observations.

\paragraph{Interpretation of double-ellipsoid model results.} It is physically meaningful that the parameters for APF's Gaussian ellipsoid that minimize $L^2$ error (and produce $\theta_c \approx 46^{\circ}$) relegate APF essentially to an ultra-thin layer near the upper, free interface. First, this evokes natural comparison with previous discussion that APF may be the manifestation in the amorphous bulk of near-surface atomic redistribution as has been studied within the crater function family of models and elsewhere \cite{kalyanasundaram-etal-APL-2008,kalyanasundaram-etal-JPCM-2009,norris-etal-2009-JPCM,norris-etal-NIMB-2013}. Second, while the displacements within the ion-implanted ellipsoid are expected to generate isotropic stresses, it is possible that the collisions above that region, higher in the nuclear collision cascade, are directional enough to result in deviatoric stresses after rearrangements.

\paragraph{1-20keV range for Ar$^+ \to$ Si.} We have suggested that two minimal conditions for a successful theory of ion-induced nanopatterning are the correct prediction of $\theta_c$ and the correct prediction of angle-dependent ripple wavelengths. An interesting aspect of the $\theta_c$ selection problem is that for some systems, patterns are strongly suppressed within certain energetic ranges. Such a phenomenon is observed for the Ar$^+ \to$ Si system. At energies as low as 30eV, ripple patterns form at $\theta_c \approx 55^{\circ}$, while for 250eV-1keV, patterns form for $\theta_c \approx 45-48^{\circ}$. However, for about 1-20keV Ar$^+ \to$ Si, patterns do not appear at any angle of incidence up to $\theta = 75^{\circ}$, suggesting that either $\theta_c \geq 75^{\circ}$ or that patterns are completely suppressed for any $\theta$ within this energy range. Then, although experimental data is sparse, patterns seemingly begin to reappear at 20keV and are present through at least 100keV \cite{hofsass-bobes-zhang-JAP-2016}, with $\theta_c$ apparently as low as $30^{\circ}$ in some cases \cite{carter-etal-REDS-1977}, albeit for high fluences. Hence a difficulty in $\theta_c$ prediction is in accounting for this peculiar pattern of energy dependence.

The double-ellipsoid model has suggested that the most parsimonious explanation of in-plane stress and $\theta_c$ data for 250eV Ar$^+ \to$ Si involves modeling APF as existing within an ultra-thin layer near the upper interface. We speculative that it is possible that \textit{even in the absence of a melt-cycle} enough anisotropy is retained in the displacement field to induce deviatoric stresses well-described by the APF stress tensor. Under this assumption, anisotropic displacements driven by electronic stopping near the upper interface need to be spatially-separated from the region of strong isotropic displacements driven by nuclear stopping within the ion-implanted region in order for the anisotropic displacements to develop into the observed, low-energy APF. This is because the anisotropic displacements, if encroached upon by large numbers of isotropic displacements, will be ``drowned out" and anisotropy will be lost.  Accordingly, we may consider the ratio $\frac{\alpha}{a}$ as a marker of this ``encroachment". When $\frac{\alpha}{a}$ is large, the ion-implanted region is closer to the upper interface, and the region of isotropic displacements invades the region of hypothesized, anisotropic displacements. When $\frac{\alpha}{a}$ is small, the ion-implanted region is relatively deep within the film and the hypothesized, anisotropic displacements have room to develop into deviatoric stresses. From SRIM, we generate a curve of $\frac{\alpha}{a}$ against beam energy for Ar$^+ \to$ Si.

\begin{figure}[h!]
	\centering
	\includegraphics[totalheight=5cm]{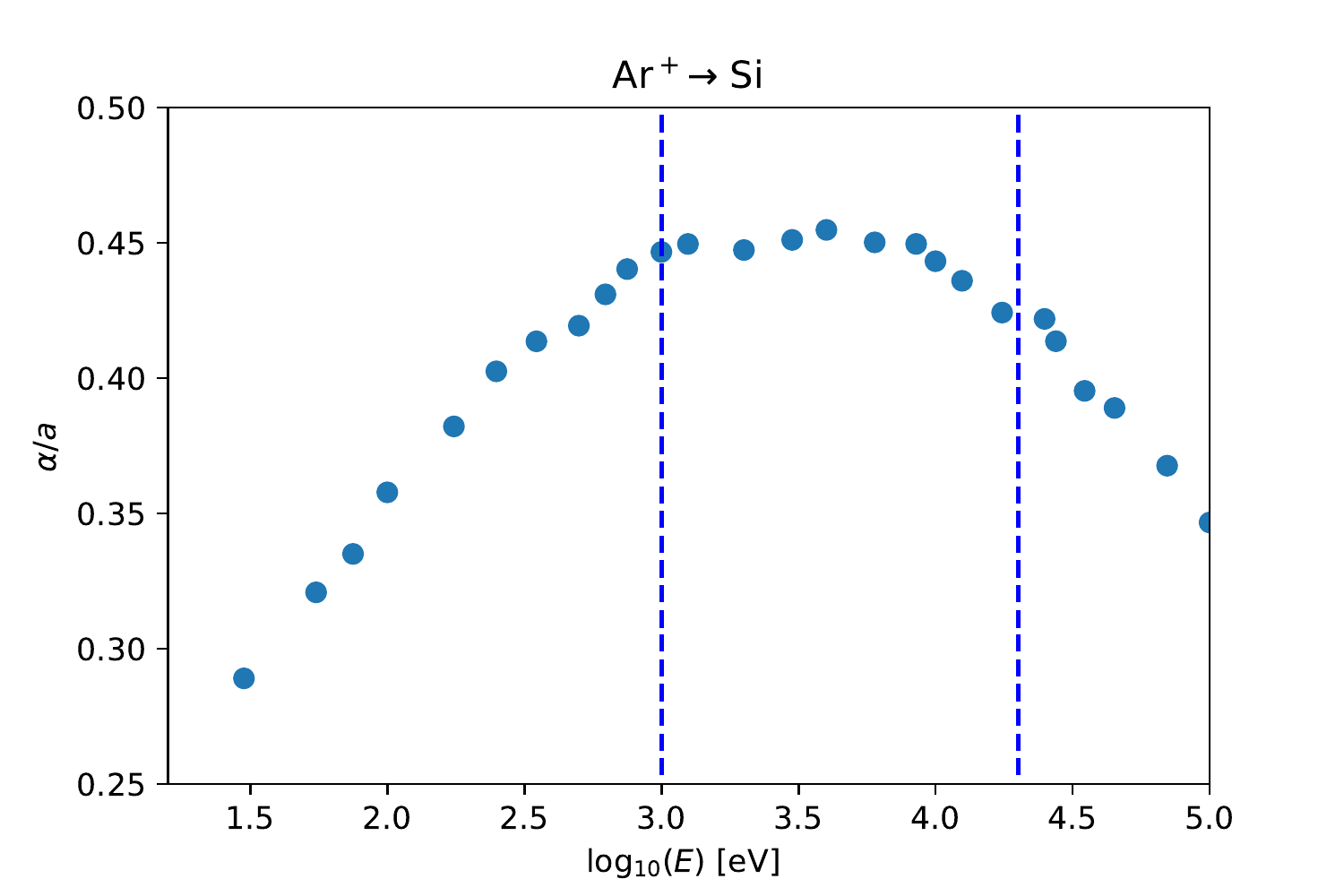}
	\includegraphics[totalheight=5cm]{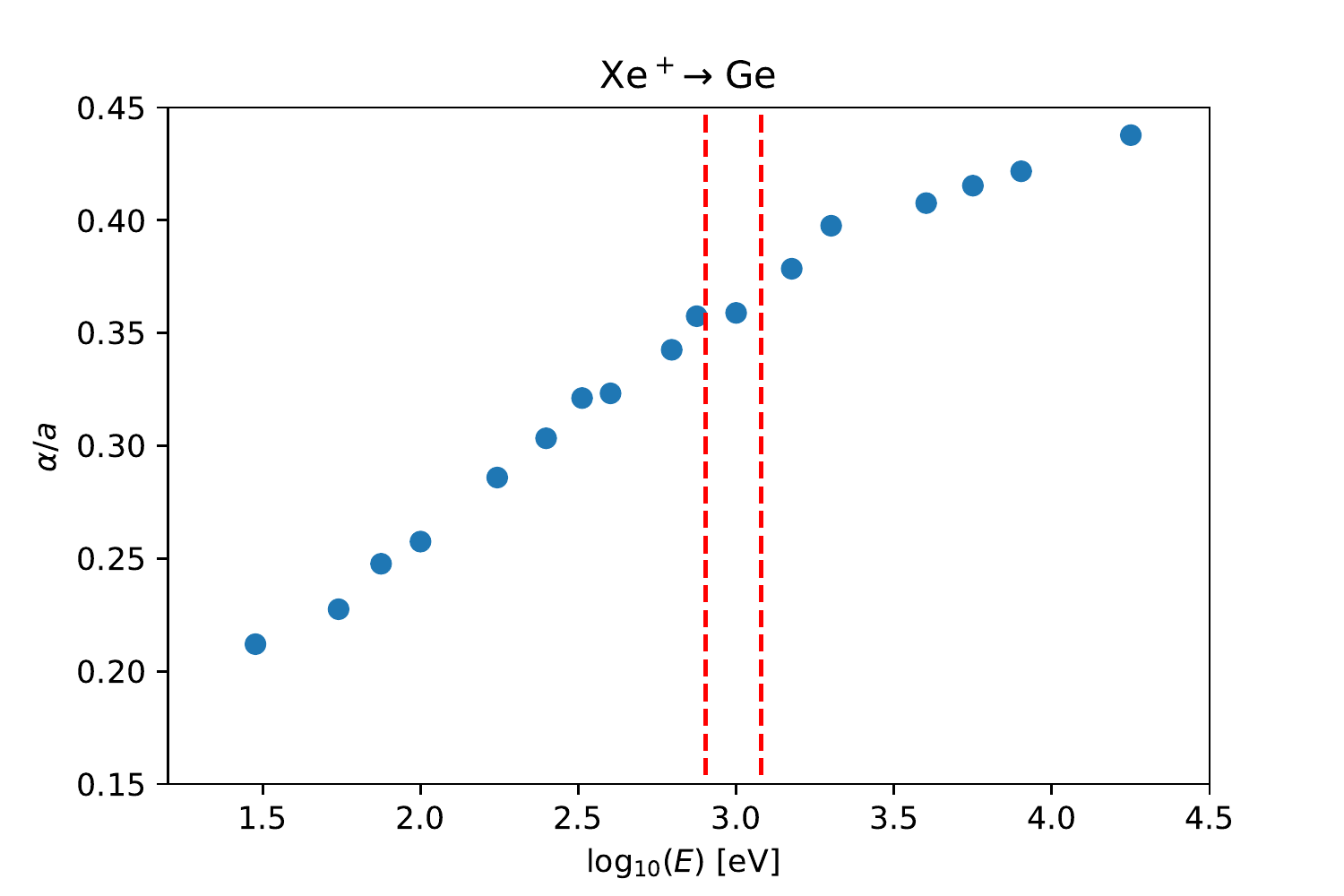}	
	\caption{\textbf{Left:} The ratio $\frac{\alpha}{a}$ against beam energy for Ar$^+ \to$ Si irradiation as computed from SRIM. \textbf{Dashed blue lines} mark the 1keV-20keV region where patterns are strongly suppressed \cite{hofsass-bobes-zhang-JAP-2016}. Notice that this region coincides well with the peak of $\frac{\alpha}{a}$, suggesting that the hypothesized region of anisotropic displacements is encroached upon by the ion-implanted region associated with many local isotropic rearrangements, leading to the anisotropy being ``drowned out". \textbf{Right:} The ratio $\frac{\alpha}{a}$ against beam energy for Xe$^+ \to$ Ge irradiation as computed from SRIM. The \textbf{dashed red lines} mark the energetic region at which patterns are suppressed at $65^{\circ}$ and later return for $75^{\circ}$ \cite{Teichmann2013}. We are aware of no experimental data for $\theta_c$ in the literature for Xe$^+ \to$ Ge in the 2keV-100keV range.}
\end{figure}

We find that the energetic range from 1-20keV, where patterns are strongly suppressed for Ar$^+ \to$ Si, coincides quite well with the peak of the ratio $\frac{\alpha}{a}$ as a function of energy. This qualitatively agrees with our hypothesized description of APF in the nuclear stopping regime and lends further weight to the hypothesis of spatially-separated, competing mechanisms.

\paragraph{Xe$^+ \to$ Ge in the 400eV-2keV range.} Teichmann et al \cite{Teichmann2013} have conducted experiments for Xe$^+ \to$ Ge at 65$^{\circ}$ and 75$^{\circ}$ for 400eV, 800eV, 1.2keV and 2keV. It was observed that patterns form for irradiation of Ge by Xe$^+$ for $\theta_c \approx$ 65$^{\circ}$ at 400eV and 800eV. However, for 1.2keV, patterns have faded and by 2keV they have disappeared altogether. Increasing the irradiation angle to 75 degrees results in the reappearance of patterns at 1.2keV and 2keV. We hypothesize that this energetic dependence may be driven by the same phenomenon as that in Ar$^+ \to$ Si described above. Again, we find that the energetic region associated with weakened pattern formation coincides with encroachment of the ion-implanted region into the ultra-thin region containing the hypothesized anisotropic displacements. We are aware of no experimental data for angle-dependent pattern formation above 2keV for Xe$^+ \to$ Ge, but such data would be of interest. We hypothesize that patterns suppression for Xe$^+ \to$ Ge would strengthen with increasing energy, as Xe$^+ \to$ Ge does not exhibit the same ``peak" as Ar$^+ \to$ Si.

\paragraph{Ne$^+ \to$ Ge, 400eV-2keV.} We note that \cite{Teichmann2013} have also observed no pattern formation for Ne$^+ \to$ Ge for any angle of incidence up to $75^{\circ}$ under 400eV-2keV irradiation. Indeed, as with the above cases where patterns are strongly suppressed, the ion-implanted region is ``high up" in the film. Under our hypothesis, this would tend to ``drown out" the localized anisotropic displacement fields (perhaps analogous to shear transformation zones \cite{vandillen-etal-prb-2006,wang-etal-APL-2012-metallicglasses,egami-etal-metals-2013}) by saturating them with localized \textit{isotropic} displacement fields. For 400eV, we have $\frac{\alpha}{a} \approx .601$, for 800eV we have $\frac{\alpha}{a} \approx .605$, for 1.2eV $\frac{\alpha}{a} \approx .614$ and for 2keV  $\frac{\alpha}{a} \approx .596$. These ratios $\frac{\alpha}{a}$ are, indeed, larger than those seen to suppress pattern formation in Ar$^+ \to$ Si, and may be interpreted as a significant amount of encroachment of local isotropic displacement fields into near-surface region within which APF may develop. While we have not established \textit{causality}, comparison with these three experimental systems has established \textit{consistency}: something qualitatively similar to what we have suggested does appear to account for trends. Exploration of this analysis and these observations, we hope, will lead to the development of constitutive relations for ion-irradiated amorphous matter that rely less on phenomenology and are more closely linked to the underlying physics. 

\section{Discussion}
\subsection{Implications for hydrodynamic-type theory}
\paragraph{Improved physical characterization.} In the present work, we have significantly advanced the hydrodynamic, stress-based theory of nano-scale pattern formation in irradiated, amorphizable materials. Analogous to the Bradley-Harper theory, which connected Sigmund's rotating-ellipse model of power deposition to \textit{surface} evolution, we have now connected the same rotating-ellipse model to the \textit{bulk dynamics} as well as the determination of the exact shape of the lower interface as adiabatically coupled to the upper within a horizontal and vertical shift. These developments have enabled a careful examination of the two apparent mechanisms operating within the film bulk: an anisotropic plastic flow and an isotropic swelling. Due to the increased detail of our theoretical approach, we have obtained improved estimates of parameter values within about an order of magnitude of other parameter estimates for the same, or similar, systems. 

The theoretical framework and closed-form expressions presented here will be easily-combined with future experimental measurements of angle-dependent stress, such as those of \cite{perkinsonthesis2017}, in order to develop stress-based models of ion-irradiated thin films with increasing quantitative accuracy and predictive power. Particularly, our detailed treatment of the interface relation will immediately improve the accuracy of future wafer curvature stress measurements, and furthermore eliminates a degree of freedom which appears to have previously contributed to the delayed convergence of the scientific community upon a unified theory. Our treatment of power deposition in the bulk is also easily-understood and readily applicable to other linear stability studies of similar systems, and serves as the bulk-analogue of the study of \cite{bradley-PRB-2011b}.

\paragraph{Modeling per-mechanism spatial variation.} We have determined that the hypothesis that both anisotropic plastic flow and isotropic swelling originate within the collision cascade is strongly incompatible with the experimental data available to us, a finding which would have been inaccessible in the absence of our spatially-resolved analysis. This is, at first, perplexing. However, we have considered the possibility that while isotropic swelling originates in the collision cascade (in qualitative agreement with other work), anisotropic plastic flow may originate elsewhere and does not ``share space". Minimizing $L^2$ error between our model and experimental data over a wide range of possible distributions for anisotropic plastic flow places APF at \textit{the very top of the film}. This leads us to an alternative hypothesis: that anisotropic plastic flow is primarily a \textit{near-surface phenomenon}, originating ``up-beam". This hypothesis agrees very favorably with discussion in \cite{norris-PRB-2012-linear-viscous,NorrisAziz_predictivemodel} that APF may share a common origin with near-surface redistribution as studied within the crater function framework; with the scaling of the APF stress tensor in the nuclear stopping regime documented by \cite{van-dillen-etal-APL-2001-colloidal-ellipsoids,van-dillen-etal-APL-2003-colloidal-ellipsoids,van-dillen-etal-PRB-2005-viscoelastic-model}; and with observations of pattern suppression for Ar$^+$ on Si for $\geq$ 1.2keV as considered in \cite{hofsass-bobes-zhang-JAP-2016}, as well as those for Xe$^+ \to$ Ge for 400eV-1.2keV in \cite{Teichmann2013}. In the latter two systems, it appears that suppression of patterns is associated with energy levels for which the nuclear collision cascade is relatively high-up in the amorphous bulk. In such cases, we hypothesize that isotropic displacements ``drown out" the anisotropic displacements near the upper interface, creating a strong stabilization effect greater than that of defect production alone. That is: it may be that the ion-implanted region is the locus of isotropic swelling, which is itself stabilizing, but that isotropically-swollen regions may actually competitively-inhibit the development of APF when the two regions are in close-enough contact, leading to an \textit{additional} stabilization effect for certain collision cascade geometries. This nonlinearity would account for the apparently sudden, bifurcation-like suppression of patterns for about 1-20keV in Ar$^+ \to$ Si, and the return of patterns above about 20keV. This hypothesis should, in principle, be verifiable through MD simulations and wafer irradiation experiments.

\paragraph{A possible bridge between the nuclear and electronic stopping regimes.} Fascinatingly, when we have a careful treatment of the lower interface and make \textit{any} assumption about the physical origin of APF, an apparent lower bound for $\theta_c$ for $\frac{A_I}{A_D} \to 0$ is $\sim 30^{\circ}$. Physically, this is interesting because it appears to align with the experimentally-observed lower bound seen in studies of irradiation of Si by noble gas ions \cite{carter-etal-REDS-1977,carter-JAP-1999,hofsass-bobes-zhang-JAP-2016}. It is noteworthy that the void formation and defect dynamics that are suspected to drive isotropic swelling within the nuclear collision cascade are sensitive to temperature, with high temperatures tending to anneal away defective regions, forcing local recrystallization of the lattice. We might therefore expect the melt-cycle at high energies, a well-established physical origin of APF in the electronic stopping regime, to compete with IIS for space within the film, eventually dominating IIS sufficiently that $\theta_c \approx 30^{\circ}$ is observed. If so, it may be possible to directly bridge the nuclear and electronic stopping regimes, creating a single, unified theory that encompasses all energies from that of negligible sputtering well into the MeV range. This prospect is made all the more likely by our observations that APF is most parsimoniously assigned to an ultra-thin region near the upper interface where elastic collisions at the beginning of the nuclear collision cascade, below a certain electronic stopping power, may still induce anisotropic displacements even if there is not enough electronic power deposition to induce a proper melt cycle along the ion track.

\subsection{Open questions and future directions}
\paragraph{Per-mechanism spatial variation.} We have suggested the possibility that anisotropic plastic flow is a near-surface phenomenon, as the spatial-separation hypothesis leads to improved predictive accuracy against the hypothesis of coexisting mechanisms. We have also suggested that isotropic swelling may, indeed, be linked primarily to the collision cascade parameters which may be easily obtained through BCA simulations such as SRIM. If this is true, the next natural question is that of how to determine the correct depth-dependence of anisotropic plastic flow, as distinct from isotropic swelling. Another high-value goal would be the connection of parameter values directly to a first-principles theoretical framework, preferring that such a framework should operate in terms of parameters determinable via BCA.

One possibility, already discussed elsewhere, is that anisotropic plastic flow is the manifestation in the bulk of redistribution along the upper interface, as in the Carter-Vishnyakov model. We note also that the Carter-Vishnyakov model of prompt-regime redistribution \cite{carter-vishnyakov-PRB-1996} assumes a perfectly spherical collision cascade as an order-of-magnitude estimate and is confined to the upper interface, making no explicit consideration of the bulk; hence an alternative is that Carter-Vishnyakov restribution is a special case of the \textit{near}-surface phenomenon we have described as anisotropic plastic flow. The assumption of redistribution \textit{at} the surface rather than \textit{near} the surface may also account for some of the disagreements seen between theory and experiment as in, e.g., PyCraters \cite{norris-arXiv-2014-pycraters} with respect to $\theta_c$ selection. It is noteworthy that the predictions of PyCraters are worse when compared to experimental values for ion-target-energy combinations with deeper collision cascades, suggesting the relevance of the bulk physics and the need to spatially resolve these mechanisms.

Because of this, we propose that a high research priority should be the theoretical unification and combined treatment of models restricted to the surface, traditionally comprising erosion, redistribution, and possibly surface diffusion (in high-temperature systems), with the more recent class of hydrodynamic, stress-based models focused on the bulk, which at various times included effective body-forces or analogues thereof, anisotropic plastic flow, surface-confined viscous flow, and isotropic swelling. This is necessary in order to avoid the possibility of ``double-counting" mechanisms which may be, in reality, the same phenomenon appearing in two different places due not to the underlying physics but, rather, due to the modeling approaches taken. An unintended, but welcome, consequence here would be strengthened confidence in parameter estimates for \textit{all} such mechanisms due to simultaneous fitting, as in \cite{norris-etal-SREP-2017}.

A natural approach to attempting to resolve this question would be a data-driven comparison of experimental critical angle observations, the present theoretical framework, and the treatment of the location of anisotropic plastic flow as a function of some data set. With a large enough data set, we could then attempt to simultaneously deduce swelling rates and the depth-dependence shape of anisotropic plastic flow, possibly as four or five fit parameters in an assumed, plausible functional form against a large data set. A sufficiently low relative error alongside consistent trends would serve as validation of the hypothesis and may lead to further theoretical developments, eventually culminating in a fully predictive, first-principles theory.

\paragraph{Other stabilizing mechanisms.} At the same time, the possibility exists that the apparent disagreement between experimental data and the ``coexisting mechanisms" hypothesis is due not to spatially separated mechanisms, but rather some as-yet unidentified stabilizing mechanism, which coincidentally scales with energy in such a way as to explain the trends. Clearly, this cannot be ruled out until more work has been done. We remark that although we have made an effort to improve the physical accuracy of the class of hydrodynamic models here, including the treatment of the lower interface, there is one key aspect which we have neglected from our analysis. While existing hydrodynamic models borrow ideas from traditional continuum mechanics, including the no-slip and no-penetration conditions typically associated with continuum fluid flow, we note that for systems such as 250eV Ar$^+ \to$ Si, the film itself may be only a few tens of atoms thick, which strains the continuum hypothesis. Under such conditions, wherein the Knudsen number is relatively high due to large displacements versus characteristic film thickness, it is possible that \textit{slip-flow} occurs. The hypothesis that slip-flow may occur is bolstered by our finding that the shape of the lower-interface agrees strongly with the 95\%-deposition level curve of the Gaussian ellipsoid: it would be, in fact, unphysical to assert that there is \textit{no} contribution to the velocity field at the lower boundary due to the power deposited there. As we have seen that even seemingly ``insignificant" physical details may have large ramifications for the overall bulk physics, the loosening of the ``no-slip" condition may be warranted, especially if it can be reliably coupled to bulk power deposition. Here, we note that the inclusion of a scaling factor by the nominal beam energy, $E_{nom}$, in our expression of $E_D$ factors out and cannot affect $\theta_c$ in the small cross-terms limit; studying slip-flow, or other boundary conditions, might prevent the factoring-out of this constant due to the addition of new terms in the dispersion relation, which would then allow for a more direct influence of beam energy in our model, which the collision cascade statistics only indirectly impart. 

Additionally, it is well-understood that the amorphization of Si, itself, induces local changes in volume, as a-Si is less dense than c-Si, while mass is conserved; at a boundary undergoing continuous amorphization in the translating frame, this is known to induce a normal-incidence velocity field into the bulk sometimes referred to as \textit{volume convection} within the solidification theory literature (\cite{davis-solidification-book-2001} and many others). We have already seen in \cite{Swenson_2018,evans-norris-JPCM-2022} that vertical flow fields may be associated with stabilization of the upper interface against perturbations if they create a tendency for valleys to be ``filled in", and it is possible that a similar effect may be seen when the ongoing amorphization of the boundary is considered (and possibly for analogous reasons, given the isotropic character of the stress due to both mechanisms, differing only by origin). This will also allow us to distinguish between swelling at the lower boundary due to instantaneous amorphization and post-amorphization swelling in the bulk. Such a distinction could eventually become crucial for the study of materials which swell strongly upon ion-irradiation, such as germanium, and lead to refined parameter estimates in the style of \cite{norris-etal-SREP-2017}.

\paragraph{The need for more angle-dependent stress measurements.} As we have begun to incorporate more detail into our physical modeling, the need for more experimental work has increased. The wafer-curvature experiments of Perkinson \cite{perkinsonthesis2017} have been extremely valuable in the formulation of the ``separated mechanisms" hypothesis and for the purposes of comparing our own parameter estimates with others that have studied the same, or similar, systems. Other studies on different ion-target-energy combinations in the vein of Perkinson \cite{perkinsonthesis2017} would be invaluable in further developing the theory and connecting stress-based models to, perhaps, parameter values easily obtained through BCA simulations, eventually leading to a fully predictive theory.

\section*{Acknowledgments}
We gratefully acknowledge support from the National Science Foundation through DMS-1840260.

\appendix

\section{Full-spectrum dispersion relations}
\paragraph{Review of isotropic swelling result}
For completeness, we note the result from \cite{Swenson_2018},
\begin{equation}
	\sigma_{IIS} = fA_I\left( (1-\frac{\cosh(Q)+Q\sinh(Q)}{Q^2 + \cosh^2(Q)} )\frac{g_1}{h_1} - \frac{Q^2}{Q^2+\cosh^2(Q)}\right),
\end{equation}
for isotropic swelling with arbitrary interfaces and uniform strength. Here, $Q = kh_0(\theta)$.

\paragraph{Generalization of anisotropic plastic flow}
Proceeding from the work shown in the Appendix of \cite{norris-PRB-2012-linear-viscous} and taking the crystalline-amorphous interface to be arbitrary $z=g_{1}(x,y)$ instead of the vertical-translation assumption $z=h_{1}(x,y)$, we may ``recycle" almost everything. This is greatly facilitated by the appearance of the lower interface in exactly one location: the no-slip and no-penetration interface conditions, which have been combined into a single expression. We will briefly summarize the linearization, which can be read in full detail in \cite{norris-PRB-2012-linear-viscous}, and then provide the (admittedly slight) extension to it.

For the convenience of the reader, we summarize the final stages of the analysis verbatim before deviating from them at a point that we will indicate. Following linearization, we have conservation of momentum and mass yield, respectively
\begin{equation}
	\begin{gathered}
		-ik_{1}p_{1}(z) + \eta\{-(k_1^2+k_2^2)u_1 + u_1''\} = 0 \\
		-ik_{2}p_{1}(z) + \eta\{-(k_1^2+k_2^2)v_1 + v_1''\} = 0 \\
		p_{1}'(z) + \eta\{-(k_1^2+k_2^2)w_1 + w_1''\} = 0 \\
		ik_1u_{1} + ik_2v_{1} + w_{1}' = 0
	\end{gathered}
\end{equation}
whose solutions are, in general,
\begin{equation}
	\begin{gathered}
		p_{1}(z) = -2\eta[(RH + ik_1C + ik_2E)\cosh(Rz) + (RG + ik_1D + ik_2F)\sinh(Rz)] \\
		u_{1}(z) = C\cosh(Rz) + D\sinh(Rz) - \frac{ik_1}{R}[(RG + ik_1D + ik_2F)z\cosh(Rz) + (RH + ik_{1} + ik_{2}E)z\sinh(Rz)] \\
		v_{1}(z) = E\cosh(Rz) + F\sinh(Rz) - \frac{ik_2}{R}[(RG + ik_1D + ik_2F)z\cosh(Rz) + (RH + ik_{1} + ik_{2}E)z\sinh(Rz)] \\
		w_{1}(z) = G\cosh(Rz) + H\sinh(Rz) - [(RG + ik_1D + ik_2F)z\sinh(Rz) + (RH + ik_{1} + ik_{2}E)z\cosh(Rz)]
	\end{gathered}
\end{equation}
where $R = \sqrt{k_{1}^2+k_{2}^2}$ and C,D,E,F,G,H are integration constants.
Applying the combined no-slip and no-penetration condition at the lower interface yields
\begin{equation}
	\textbf{v}_{1}(x,y,0) + \frac{\partial \textbf{v}_{0}}{\partial z}(x,y,0)g_{1} = 0,
\end{equation}
implying
\begin{equation}
	C = -3fA\sin(2\theta)g_{1}(x,y), \hspace{.5cm} E = 0, \hspace{.5cm} G = 0.
\end{equation}
The coefficient $C$ is the only place where we depart from the previous work of \cite{norris-PRB-2012-linear-viscous}; both $E$ and $G$ remain zero. The influence of the arbitrary lower interface $g_{1}$ will pass into the pressure and velocity fields, and, from there, into the stress balance at the upper interface and the kinematic condition, at which point it will appear in the dispersion relation. With this modification, we consider the stress balance at the upper interface. That is,
\begin{equation}
	\textbf{T}\cdot \hat{n} = -\gamma \kappa \hat{n}
\end{equation}
at $z = h_{0} + \epsilon h_{1}(x,y)$. Although we intend to expand in the normal modes later, we will suppress this expansion for $h_{1}(x,y)$ and $g_{1}(x,y)$ until the end; this delay will prove useful. To facilitate the calculation, we first \textit{formally} linearize the above equation, leading to
\begin{equation}
	\textbf{T}_{0}\cdot \hat{n}_{1} + \textbf{T}_{1}\cdot \hat{n}_{0} = -\gamma \kappa_{1} \hat{n}_{0},
\end{equation}
at $z = h_{0}$. Here, we have
\begin{equation}
	\begin{gathered}
		\kappa_{1} = -(h_{1,yy} + h_{1,xx}); \\
		\hat{n}_{0} = <0, 0, 1>; \\
		\hat{n}_{1} = <-h_{1,x}, -h_{1,y}, 0>; \\
		\textbf{T}_{0} = -6\eta fA
		\begin{bmatrix} 
			\cos(2\theta) & 0 & 0 \\
			0 & \cos^2(\theta) & 0 \\
			0 & 0 & 0 \\
		\end{bmatrix}; \\
		\textbf{T}_{1} =
		\begin{bmatrix} 
			-p_{1} + 2\eta u_{1,x} & \eta (u_{1,y}+v_{1,x}) & \eta (u_{1,z} + w_{1,x}) \\
			\eta(v_{1,x} + u_{1,y}) & -p_{1} + 2\eta v_{1,y} & \eta (v_{1,z} + w_{1,y}) \\
			\eta(w_{1,x}+u_{1,z}) & \eta(w_{1,y}+v_{1,z}) & -p_{1} + 2\eta w_{1,z} \\
		\end{bmatrix}.
	\end{gathered}
\end{equation}
We will ``recycle" much of the previous work by making a sequence of useful observations: $\textbf{T}_{1}\cdot \hat{n}_{0}$ merely selects the third column of $\textbf{T}_{1}$, and within this column, the coefficient $C$ is never differentiated with respect to $x$ or $y$, because $C$ appears only in $u_{1}$ and $v_{1}$. Then the $\bar{\alpha}, \bar{\beta}, \bar{\gamma}$ terms, (A13) in \cite{norris-PRB-2012-linear-viscous}, are unchanged except for $h_{1} \to g_{1}$ in the term involving $\sin(2\theta)$ on the third column vector. These constants serve the same role in the solution as originally, with coefficients $D,F$ and $H$ solved in terms of them. Ultimately, $D,F$ and $H$ alter $w_{1}(h_{0})$, but $u_{0}(h_{0})$ and $v_{0}(h_{0})$ are, of course, unchanged. Then we construct a matrix equation for the stress balance at the upper interface as
\begin{equation}
	\begin{gathered}
		\begin{bmatrix}
			R\bar{C} + \frac{k^2_{1}}{R}(\bar{C} + 2Q\bar{S}) && \frac{k_1k_2}{R}(\bar{C}+2Q\bar{S}) && -2ik_1Q\bar{C} \\
			\frac{k_1k_2}{R}(\bar{C}+2Q\bar{S}) && R\bar{C} + \frac{k^2_{2}}{R}(\bar{C} + 2Q\bar{S})&& -2ik_2Q\bar{C}\\ 
			-2ik_1Q\bar{C} && -2ik_2Q\bar{C}&& 2R(\bar{C} - Q\bar{S})
		\end{bmatrix}
		\begin{bmatrix}
			D \\
			F \\
			H\\
		\end{bmatrix}
		=
		\begin{bmatrix}
			\bar{\alpha} \\
			\bar{\beta} \\
			\bar{\gamma} \\
		\end{bmatrix}
	\end{gathered},
\end{equation}
where $Q=h_{0}R, \bar{C} = \cosh(Q), \bar{S} = \sinh(Q)$, and
\begin{equation}
	\begin{bmatrix}
		\bar{\alpha} \\
		\bar{\beta} \\
		\bar{\gamma} \\
	\end{bmatrix}
	= -\gamma h_{1}
	\begin{bmatrix}
		0 \\
		0 \\
		R^2
	\end{bmatrix}
	+
	6\eta f A h_{1}
	\begin{bmatrix}
		ik_{1}\cos(2\theta) \\
		ik_{2}\cos^2(\theta) \\
		0
	\end{bmatrix}
	+
	3\eta f A g_{1}\sin(2\theta)
	\begin{bmatrix}
		R\bar{S} + \frac{k_{1}^2}{R}(\bar{S} + 2Q\bar{C})  \\
		\frac{k_1k_2}{R}(\bar{S} +2Q\bar{C}) \\
		-2ik_{1}Q\sinh(Q)
	\end{bmatrix}
\end{equation}
from which we may solve for D,F and H. From the kinematic condition at the upper interface, linearized as
\begin{equation}
	h_{1,t} = w_{1}(h_{0}) - u_{0}(h_{0})h_{1,x} - v_{0}(h_{0})h_{1,y},
\end{equation}
where
\begin{equation}
	\begin{gathered}
		u_{0}(h_{0}) = 3fA\sin(2\theta)h_{0}; \\
		v_{0}(h_{0}) = 0, \\
		w_{1}(h_{0}) = -6fA \tilde{h}_{1} \frac{\cos(2\theta)(h_0k_1)^2 + \cos^2(\theta)(h_0k_2)^2}{1+2Q^2+\cosh(2Q)} \\
		- 3fA\tilde{g}_{1}\sin(2\theta)(ih_0k_1)
		\left( \frac{2 \cosh(Q)[Q^2 + \sinh^2(Q) ]}{1+2Q^2+\cosh(2Q)} - \cosh(Q) \right) \\
		-\frac{\gamma \tilde{h}_{1}}{2\eta h_{0}} \frac{Q(\sinh(2Q) - 2Q)}{1+2Q^2+\cosh(2Q)}
	\end{gathered}
\end{equation}
we obtain the dispersion relation, which will change only very subtly. This is quite intuitive when we note that the only term that can possibly change is the $w_{1}$ term, because the $u_{0}$ and $v_{0}$ terms were determined prior to the perturbation of the free interface, where the steady state for $h$ is $h_{0}$ and the steady state for $g$ is 0. Due to having retained general $g_{1}(x,y)$, some of the coefficients of terms in the expression for $w_{1}$ will now have $g_{1}$ where they originally had $h_{1}$. In order to isolate dispersion relation $\sigma$, divide through by $\tilde{h}_{1}$, we obtain the generalization of the original result,
\begin{equation}
	\begin{gathered} 
		\sigma(k_1,k_2) = -6fA \frac{\cos(2\theta)(h_0k_1)^2 + \cos^2(\theta)(h_0k_2)^2}{1+2Q^2+\cosh(2Q)}
		- 3fA\sin(2\theta)(ih_0k_1) \\
		- 3fA\sin(2\theta)(ih_0k_1)\times \frac{\tilde{g}_{1}}{\tilde{h}_{1}}
		\left( \frac{2 \cosh(Q)[Q^2 + \sinh^2(Q) ]}{1+2Q^2+\cosh(2Q)} - \cosh(Q) \right) \\
		-\frac{\gamma}{2\eta h_{0}} \frac{Q(\sinh(2Q) - 2Q)}{1+2Q^2+\cosh(2Q)}
	\end{gathered}
\end{equation}
where $\frac{g_1}{h_{1}}$ is the ratio of the expansions of the perturbed lower and upper interfaces in normal modes, as in the main text. We note the agreement with the physical interpretation given in the original analysis of \cite{norris-PRB-2012-linear-viscous}, which describes the third term as associated with the effect of the nonplanar lower interface under the influence of the beam.
\printbibliography

\end{document}